\documentclass[12pt,english]{article}

\makeatletter
\newcommand*{\rom}[1]{\expandafter\@slowromancap\romannumeral #1@}
\makeatother

\usepackage[latin9]{inputenc}
\usepackage{geometry}
\usepackage{verbatim}
\usepackage{prettyref}
\usepackage{amsmath}
\usepackage{amssymb}
\usepackage{enumitem}
\usepackage{cases}
\usepackage{ytableau}
\usepackage{graphicx}
\usepackage{caption}
\usepackage{subcaption}
\usepackage{tikz}
\usepackage{mwe}

 \usepackage{slashed}
\usepackage{esvect}
\usepackage{dsfont}

\usepackage{color}
\definecolor{darkgreen}{rgb}{0,0.5,0}
\definecolor{darkblue}{rgb}{0,0,0.6}
\definecolor{purple}{rgb}{0.4,.2,0.7}
\usepackage[colorlinks=true,citecolor=black,linkcolor=black,urlcolor=black]{hyperref}
 	
\numberwithin{equation}{section}
\numberwithin{figure}{section}
\numberwithin{table}{section}

\DeclareMathOperator{\Tr}{Tr}

\newcommand{\rme}{\mathrm e}
\newcommand{\rmd}{\mathrm d}
\newcommand{\rmi}{\mathrm i}

\newcommand{\SO}{\mathop{\mathrm SO}}

\newcommand{\AdS}{\mathop{\mathrm {}AdS}}
\newcommand{\dS}{\mathop{\mathrm {}dS}}

\usepackage[backend=biber, sorting=none]{biblatex}
\addtocounter{page}{-1}

\DeclareFontShape{OT1}{cmr}{mx}{n}{<->cmr10}{}
\title{ \textbf{ Mass and spin for classical strings in \texorpdfstring{$dS_3$}{TEXT}}}
\author{\\
\href{mailto:k.parmentier@columbia.edu}{Klaas Parmentier}\\ \it \small Columbia University, Department of Physics}\date{}

\begin{document}
\fontseries{mx}\selectfont

\maketitle
\begin{abstract}
We demonstrate that all rigidly rotating strings with center of mass at the origin
of the $\dS_3$ static patch satisfy the Higuchi bound. This extends the observation of
Noumi et al. for the open GKP-like string to all solutions of the Larsen-Sanchez class.
We argue that strings violating the bound end up expanding towards the horizon and
provide a numerical example. Adding point masses to the open string only increases the
mass/spin ratio. For segmented strings, we write the conserved quantities, invariant under Gubser's algebraic evolution equation, in terms of discrete lightcone coordinates describing kink collisions. Randomly generated strings are found to have a tendency to escape through the horizon that is mostly determined by their energy. For rapidly rotating segmented strings with mass/spin $<$ 1, the kink collisions eventually become causally disconnected. Finally we consider the scenario of cosmic strings captured by a black hole in dS and find that horizon friction can make the strings longer.
\end{abstract}
\thispagestyle{empty}
\newpage
\tableofcontents
\newpage

\section{Introduction}

 In de Sitter space, the flat space Regge relation
 
 \begin{equation}\label{regge}
 M^2 \sim L/\alpha'
 \end{equation}
 
 between the mass $M$, angular momentum $L$ and Regge slope $\alpha'$, is expected to be modified when the strings become large and start to feel the curvature of spacetime. Indeed, combining \eqref{regge} with the Higuchi bound \cite{Higuchi:1986py} for unitary representations of $\SO(1,d+1)$ at large $L$
    
    \begin{equation}\label{higu}
    M^2 l^2>L^2,
    \end{equation}
    
 with $l$ being the de Sitter scale, led \cite{Noumi2019, Lust2019} to conclude that for $L>l^2/\alpha'$ unitarity is violated unless the spectrum is modified. That this happens seems clear, even without resorting to unitarity, because at this point, the string has a length comparable to $l$ and notices the curvature of $\dS$. In this note, since it is not clear what unitarity imposes at the level of classical strings, we set out to explore the bound \eqref{higu} by other means.
 
 We first limit ourselves to rigidly rotating strings in $\dS_3$, for which the analysis is analytically tractable and for which we can show explicitly that \eqref{higu} holds. In higher dimensions one can consider strings that lie in the equatorial plane. Such solutions of the Polyakov action with constraints are captured by an Ansatz of \cite{Larsen:1995bp, larsen1996}. We begin section \ref{sec: larsan} by giving a concise review of this solution class, which we call the Larsen-Sanchez class. We do this with a particular emphasis on the three different regimes that occur: strings with inward cusps, outward cusps or no cusps at all. We find an analytic expression for the ratio of their energy and angular momentum and verify \eqref{higu}. It turns out that $M l/ L$ is bounded below by the corresponding ratio for the rigidly rotating straight string, which was studied by \cite{Noumi2019}. That specific solution was first obtained by \cite{DeVega1996}, whose $\AdS_3$ version is known as the GKP string \cite{gubser2002}. Implications of the Higuchi bound for towers of string states in the context of inflation were discussed in \cite{Noumi2019, Lust2019, Scalisi:2019gfv}.

 The solutions with cusps are related via Wick rotation to Kruczenski's famous spiky strings in $\AdS$ \cite{Kruczenski_2005} as we discuss in appendix \ref{app:Kruc}.
 There is a vast literature on integrability results in these cases, starting with the work of Pohlmeyer \cite{pohlmeyer} who showed how a formulation of the string equations of motion as an $O(n)$ $\sigma$-model with Virasoro constraints  reduces to sinh-Gordon, cosh-Gordon or Liouville type equations, as described in  \cite{DeVega1993} as well. In $\AdS$ at least, such strings with cusps that grow close to the boundary are dual to single-trace boundary operators whose anomalous dimension grows logarithmically with spin \cite{Kruczenski_2005,Kruczenski:2004kw, Kruczenski:2006pk, jevicki2008}. In $\dS$ it was shown recently by \cite{bakas2016} how the different regimes can be mapped to band structures of the Lam\'e potential. 
 
After having discussed the Larsen-Sanchez class of solutions in conformal gauge, we approach the problem more generally via the Nambu-Goto action. There, we can take a cylindrical gauge to show that every rigidly rotating solution is indeed part of the Larsen-Sanchez class. It is moreover straightforward to prove that if \eqref{higu} is to be violated, by some other non-rigid solution, the string needs to extend partly beyond

\begin{equation}\label{stab}
    r > l/\sqrt{2}
\end{equation} 

This is not unexpected, since \eqref{stab} is the stability bound for the closed circular string \cite{DeVega1994a} whose (in)stability was analyzed in \cite{ Larsen1994} and which was semiclassically quantized in \cite{deVega:1994yz}. We discuss the behavior of the perturbations around such an unstable circular string when it expands towards the horizon, where from the point of view of the static patch observer the perturbations get exponentially frozen. 

The Nambu-Goto equations also lend themselves to numerical analysis, allowing us to give an example of a string that violates \eqref{higu} and expands towards the horizon, still consistent with the idea that \eqref{higu} applies only to those strings remaining within the static patch and do not reach the horizon in finite proper time. Looking at open strings, we find that the only rigidly rotating open string that goes through the origin is the GKP-like string. We can however imagine varying the setup by adding point masses to its ends, which turns out to always increase the value of $M l /L$. 

After that, we will use the method, as applied by \cite{Vegh:2015ska, Callebaut:2015fsa} in $\AdS_3$, of approximating the classical string worldsheet in a piecewise manner by totally geodesic $\dS_2 \subset \dS_3$ segments. Such strings are exact solutions, consisting of kinks moving around and colliding with each other. This can be described in terms of discrete lightcone coordinates determining the collision events. These events are subject to the algebraic evolution equation of Gubser  \cite{Gubser:2016wno}. We will quickly review this equation and the ideas leading to it. Afterwards, we shall derive expressions for the energy and angular of the segmented string that are indeed invariant under this evolution. We will show how these discrete expressions can be obtained from the Nambu-Goto conserved currents. 

The discretization makes it easier for the computer to handle the string evolution without numerical errors and allows us to further investigate how the total energy and angular momentum influence the general behavior of the string and its tendency to escape through the horizon. Typically, some time after parts of the string have left the static patch, the kinks end up in different causal patches.  At that point the algebraic evolution law breaks down and the string keeps expanding without generating new kink collisions. 

We will first generate several random strings, see Figure \ref{fig:scatters}, confirming that the general stability is mostly determined by the total energy. After that, we choose specific initial conditions with a large number of kinks, resulting in a segmented string with $M l> L$. Whereas the numerical evolution of the Nambu-Goto solution quickly becomes imprecise, here we can very accurately follow how the string rotates rapidly and expands, as in Figure \ref{fig:exit}. Working in embedding space, we see how at some point after leaving the static patch no further collisions are generated and the string expands indefinitely.

Finally we apply our results to a string which is captured by a black hole in asymptotically $\dS$-space. In recent work \cite{Xing:2020ecz}, based on \cite{Lonsdale:1988xd, Frolov:1996xw}, a cosmic string \cite{Vilenkin:1984ib} captured by a massive compact black hole is considered in flat space. Several effects such as the ejection of daughter loops and shrinking due to horizon friction were identified. Here we note that in $\dS$, due to the shape of the spectrum of the GKP-like string, the horizon friction, which lowers energy and angular momentum, can increase the length of the string. For a rotating black hole, \cite{Xing:2020ecz} demonstrate curve lengthening as well as a black hole bomb mechanism  \cite{Press:1972zz} due to circularly polarized tension waves on the captured string loop. Given our previous observations that large angular momentum makes the string expand towards the horizon, we expect this to provide a natural ending to the black hole bomb process in $\dS$.\\

{\it Note added:} At the final stages of this project, we became aware of the recent work \cite{Kato:2021rdz} which has overlap with our section \ref{sec: larsan}.
    
\section{Larsen-Sanchez strings in \texorpdfstring{$dS_3$}{TEXT}}\label{sec: larsan}

At this point we are interested in rigidly rotating strings in the static patch of $\dS_3$, for which we will demonstrate \eqref{higu}. Looking to solve the Polyakov equations of motion with constraints,  we follow \cite{larsen1996} and review their Ansatz in static patch coordinates, where the metric $g_{\mu \nu}$ is given by

\begin{equation} \label{statpatch}
    \rmd s^2 = g_{\mu \nu} \rmd X^\mu \rmd X^\nu = -(1-\frac{r^2}{l^2}) \rmd t^2 + \frac{\rmd r^2}{1- \frac{r^2}{l^2}} + r^2 \rmd \varphi^2.
\end{equation}

We need to solve the Polyakov equations

\begin{equation}\label{eom}
    \partial_\alpha \partial^\alpha X^\mu = - \Gamma^{\mu}_{\nu \rho} \partial_\alpha X^\nu \partial^\alpha X^\rho.
\end{equation}

where $\Gamma^{\mu}_{\nu \rho}$ are the Christoffel symbols. The conformal flatness of the induced metric on the worldsheet in ensured by the constraints

\begin{equation}\label{conpol}
    g_{\mu \nu}\dot{X}^\mu X'^\nu = 0 = g_{\mu \nu}\dot{X}^\mu \dot{X}^\nu + g_{\mu \nu}X'^\mu X'^\nu.
\end{equation}

The Ansatz of Larsen and Sanchez is to take \cite{larsen1996}

\begin{equation}\label{ansatz}
    X^\mu = (t, r, \varphi) = (c_1 \tau +f(\sigma), r(\sigma), c_2 \tau + g(\sigma) ).
\end{equation}

where the constants $c_1, c_2$ are real and $f$, $g$ are real functions of $\sigma$. Letting $\tau$ run while keeping $\sigma$ fixed, we see that the string indeed does not change its shape, which is why we call it rigidly rotating. The equations of motion \eqref{eom} and constraints \eqref{conpol} are satisfied when

\begin{equation}\label{fgeq}
        f' = \frac{k_1}{1-\frac{r^2}{l^2}},\hspace{0.5cm} g' =\frac{ k_2}{r^2},\hspace{0.5 cm} c_1 k_1 - c_2 k_2 = 0
\end{equation}

and 

\begin{equation}\label{larsanreq}
        r'^2 = -(1-\frac{r^2}{l^2})(r^2 c^2_2+\frac{k^2_2}{r^2}) +k^2_1 + (1-\frac{r^2}{l^2})^2c^2_1,
\end{equation}

where $k_1, k_2$ are again real constants. By interchanging the direction of rotation they can be taken positive. The above Ansatz can also be used with $\sigma$ and $\tau$ interchanged. The resulting surface that is traced out in spacetime will be the same but the timelike and spacelike worldsheet coordinates are interchanged. The induced metric on the string is

\begin{equation}\label{larsaninduced}
      \rmd s^2 =  (c^2_1-(\frac{c^2_1}{l^2}+c^2_2)r^2)(-\rmd \tau^2 + \rmd \sigma^2).
\end{equation}

\subsection{Solutions and their characterization}

Special cases of \eqref{ansatz} include $c_1 = k_2=0$, which gives the oscillating closed string \cite{DeVega1994a}. The GKP-like open rotor \cite{DeVega1996} is obtained for $k_1 = k_2 =0$.  For $c_1=c_2=0$ the worldsheet degenerates to a worldline and we get a point particle instead. We will first review the explicit solutions, which did not appear in the original paper, but showed up in slightly different notation in \cite{bakas2016}.  First note that both $\tau, \sigma$ can be rescaled such that $c_1^2 + c^2_2l^2=l^2$. Then we call $c_1 = l \sin a$ and $c_2 = \cos a$ for some $a$ which we can choose between $0$ and $\pi/2$, by picking a convention for the direction of rotation. Now we can find out at which radii the string reaches an extremum, by equating \eqref{larsanreq} to zero. The result is, after using $c_1 k_1 = c_2 k_2$,

\begin{equation}
    r_a = l \sin a, \hspace{0.5 cm} r_b = \frac{l}{\sqrt{2}}(1-\sqrt{1-4  \frac{k^2_1\sin^2 a}{l^2\cos^4 a} })^{\frac12},\hspace{0.5cm}r_c = \frac{l}{\sqrt{2}}(1+\sqrt{1-4 k^2_1 \frac{k^2_1\sin^2 a}{l^2\cos^4 a}  })^{\frac12}.
\end{equation}

If $r_b$ and $r_c$ are real we can call them $r_b = l \sin{b}$ and $r_c = l \cos{b}$ for some $b \in [0, \pi/4]$ since their squares sum to 1. When they are complex conjugate one finds that the discriminant of the Weierstrass elliptic function that solves \eqref{larsanreq} is negative and that it is not satisfactory as a solution. Then, again up to orientation reversal, we can call $k_1 = \frac12 l \sin{2b} \cos{a} $ and $k_2 = \frac12 l^2 \sin{2b} \sin{a}$. We then find that a bounded solution is given by

\begin{equation}
    r = l (\mathfrak{p}[\sigma+\mathfrak{w}]+\frac16 (3-\cos{2a}))^{\frac12}
\end{equation}

with $\mathfrak{p}$ the Weierstrass elliptic function with imaginary half-period  $\mathfrak{w}$ and Weierstrass invariants\footnote{See \cite{Gradshteyn:1702455} or the appendix of \cite{bakas2016} for an exposition of the most relevant properties of the Weierstrass elliptic functions.}

\begin{equation}
    g_2 = \frac16(4 + \cos{4a}+3\cos{4b}), \hspace{0.5 cm} g_3 = \frac{1}{54}\cos{2a}(8 -\cos{4a} +9\cos{4b}).
\end{equation}

Now we can integrate \eqref{fgeq} using 5.141.5 in \cite{Gradshteyn:1702455}

\begin{equation}
    f =  l \frac{\cos{a}\sin{2b}}{2\mathfrak{p}'[v]}\log{\frac{\mathfrak{s}[\sigma+v+\mathfrak{w}]}{\mathfrak{s}[\sigma-v+\mathfrak{w}]}-2(\sigma+\mathfrak{w})\zeta[v]}
\end{equation}

where $v = \mathfrak{p}^{-1}[\frac16 (3+\cos{2a})]$, $\zeta$ is the Weierstrass $\zeta$-function and by abuse of notation, since $\sigma$ is already a worldsheet coordinate, we used $\mathfrak{s}$ to denote the Weierstrass $\sigma$-function. Similarly

\begin{equation}
    g = -  \frac{\sin{a}\sin{2b}}{2\mathfrak{p}'[w]}\log{\frac{\mathfrak{s}[\sigma+w+\mathfrak{w}]}{\mathfrak{s}[\sigma-w+\mathfrak{w}]}-2(\sigma+\mathfrak{w})\zeta[w]}
\end{equation}

where $w = \mathfrak{p}^{-1}[\frac16 (\cos{2a}-3)]$. When using the above in Mathematica, one has to be careful about branch cuts\footnote{The interpretation of the branched worldsheet as describing (in)finitely many strings at the same time, has been discussed in \cite{DeVega1994a}, there referred to as the multi-string property.}, and in particular it is easier to plot the strings using numerical integration. The Weierstrass elliptic function is periodic under a shift of $\sigma$ equal to twice the real half-period\footnote{In Mathematica this is either WeierstrassHalfPeriodW1 or -W1-W2.}. Moreover, $t$ and $\varphi$ are quasiperiodic in the sense that after one period they shift by a constant $\Delta t$ and $\Delta \varphi$ that can be obtained explicitly using the integrals in \cite{Gradshteyn:1702455}. We are interested now in closed strings, symmetric around the origin. To have a string that closes in on itself at a fixed time $t$, we need

\begin{equation}\label{closed}
   - \cot(a)\Delta t + \Delta \varphi = 2\pi \frac{m}{n}
\end{equation}

which means that after $n$ periods the string will have gone around the origin $m$ times before closing in on itself. In the next subsection we discuss some numerically obtained examples with small $m,n$. First we observe that at $r_a$ we have $r'$=0 but also

\begin{equation}
    -\cot(a) f' +g' =0, \quad \text{at }\;r_a.
\end{equation}

This means that the angular derivative at fixed time also vanishes and that there is a cusp. We can then distinguish three regimes, apart from the previously mentioned special cases, using that \eqref{larsanreq} gets positive at large $r$ and negative as $r\to 0$. When $r_a < r_b < r_c$ the cusps are on the inside of the string. They move with the speed of light. The worldsheet is of course still Lorentzian, though from \eqref{larsaninduced} we can see that in this case it is $\sigma$ that is timelike. When $r_a = r_b< r_c$ we have a point-particle moving around at the speed of light. The second regime is when $r_b < r_a < r_c$ in which case the string extends between $r_b$ and $r_a$ and has cusps on the outside. A final regime\footnote{This regime does not seem to exist in $\AdS$ since the equation for the extrema only has only 2 real roots instead of three and therefore only has the regimes with cusps on the inside or outside depending on their order. Related to this, the straightforward Wick rotation of Kruczenski's solutions gives the first two regimes as we show in \ref{app:Kruc}. For the third regime an extra shift is needed.} occurs when $r_b < r_c < r_a$. In this case there are no cusps and the string extends between $r_b$ and $r_c$. Note that $r=l/\sqrt{2}$ lies in between these two, where $r=l/\sqrt{2}$ is the radius at which a static circular string exists \cite{DeVega1994a}. At smaller radii those circular strings oscillate through the origin and at larger radii they expand towards the horizon.


\subsection{Mass and angular momentum}

Associated with the Killing vectors $\partial_t$ and $\partial_\varphi$ are conserved worldsheet currents $J^\xi$ and $L^\xi$ for the energy and angular momentum respectively. They are obtained from the Noether procedure and satisfy 

\begin{equation}
    \partial_\xi J^\xi = \partial_\xi L^\xi = 0
\end{equation}

with $\xi$ ranging over $\tau$ and $\sigma$. When integrated over the worldsheet at fixed time $t$, or over any homologous cycle, they yield the energy and angular momentum. Using \eqref{ansatz}, we obtain

\begin{equation}
    J^\tau = -(1-\frac{r^2}{l^2})l \sin{a}, \hspace{0.5 cm} J^\sigma = \frac12 l \sin{2b}\cos{a}
\end{equation}

and for the angular momentum

\begin{equation}
    L^\tau = -r^2 \cos{a}, \hspace{0.5 cm} L^\sigma = \frac12 l^2 \sin{2b}\sin{a}.
\end{equation}

Now we integrate over a closed loop, which exists at fixed $t$, but not at fixed $\tau$.  In order for $t$ to remain the same, when we integrate over the period $\Delta \sigma$, there is an accompanying change of $\Delta \tau = -\frac{\Delta t}{l \sin{a}}$. We can also write $\Delta t$ as an integral over $\sigma$ since $t' = f' = \frac12 l \sin{2b}\cos{a}/(1-r^2/l^2)$, from \eqref{ansatz}. This gives the following expressions

\begin{equation}\label{larsanMass}
    M = \int^{\Delta \sigma}_0 \rmd \sigma [l\sin{a} (1-\frac{r^2}{l^2})-\frac{ l \sin^2{2b}\cos^2{a}}{4\sin{a} (1-\frac{r^2}{l^2})}] 
\end{equation}

\begin{equation}\label{larsanAng}
    L = \int^{\Delta \sigma}_0 \rmd \sigma[r^2\cos{a}  -\frac{ l^2 \sin^2{2b}\cos{a}}{4 (1-\frac{r^2}{l^2}) }].
\end{equation}

In the previously used notation we have $\Delta \sigma = 2 n \mathfrak{v}$, where $\mathfrak{v}$ is the real Weierstrass half-period. From this, one finds a useful combination 

\begin{equation}\label{larsanpos}
    M - \frac{\cot{a}}{l}L = l \sin{a} \int^{\Delta \sigma}_0 \rmd \sigma (1- \frac{r^2}{l^2 \sin^2 a}).
\end{equation}

It is easy to see, since the string never extends beyond $l \sin{a}$, that this quantity will be positive. In $\AdS$ there is a $\coth{a}$ on the LHS instead, which is often called $\omega$ and one typically looks at the $\omega \to 1$ limit, in which case one finds an anomalous dimension that scales logarithmically with $L$ \cite{gubser2002, Kruczenski_2005}. From this one can also observe, as before, that if $\tan{a}<1$, or equivalently $r < l/\sqrt{2}$, the stronger bound $Ml > L$ follows, since the integrand itself is positive. In general, using section 5.14 of \cite{Gradshteyn:1702455}, one can find analytic expressions for \eqref{larsanMass} and \eqref{larsanAng}. In particular one obtains the exact expression

\begin{equation}
    Ml- L = n(\alpha \zeta[\mathfrak{v}]+\beta \mathfrak{v}+ \gamma\frac{4 \mathfrak{v}\zeta[v]-4v\zeta[\mathfrak{v}]}{\mathfrak{p}'[v]} )
\end{equation}

where $v = \mathfrak{p}^{-1}[\frac16 (3+\cos{2a})]$ and

\begin{equation}
\begin{split}
\alpha &= 2l^2 (\sin{a}+\cos{a})\\
\beta &= \frac{l^2}{6}(5 \sin{a}- 5 \cos{a}+\sin{3a}+\cos{3a})\\
\gamma &= \frac{l^2}{8 \sin{a}}(1-\sin{2a}+\cos{2a})\sin^2{2b}.
\end{split}
\end{equation}

\begin{figure}[!t]
    \centering
    \includegraphics[width = 0.65\linewidth]{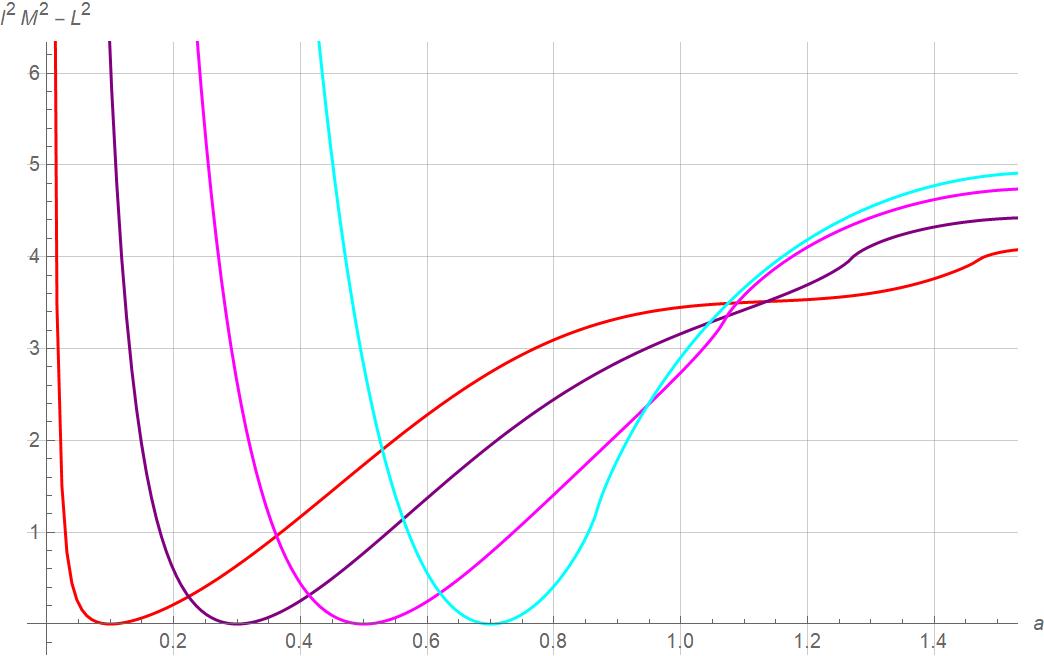}
        \caption{We plot $M^2l^2-L^2$. This quantity is seen to be positive everywhere and becomes zero at the point where the parametrization degenerates and we have a point particle moving at the speed of light. Smaller $a$ gives a string with cusps on the inside and larger $a$ gives cusps on the outside. The strange bumps occur when we go from outwards cusps to no cusps at all, namely when $a> \pi/2 - b$. Each graph corresponds to a value of $b$, starting with 0.1 in red, 0.3 in purple, 0.5 in magenta and 0.7 in cyan.}
    \label{fig: higu}
\end{figure}

\begin{figure}[!t]
    \centering
    \includegraphics[width = 0.65\linewidth]{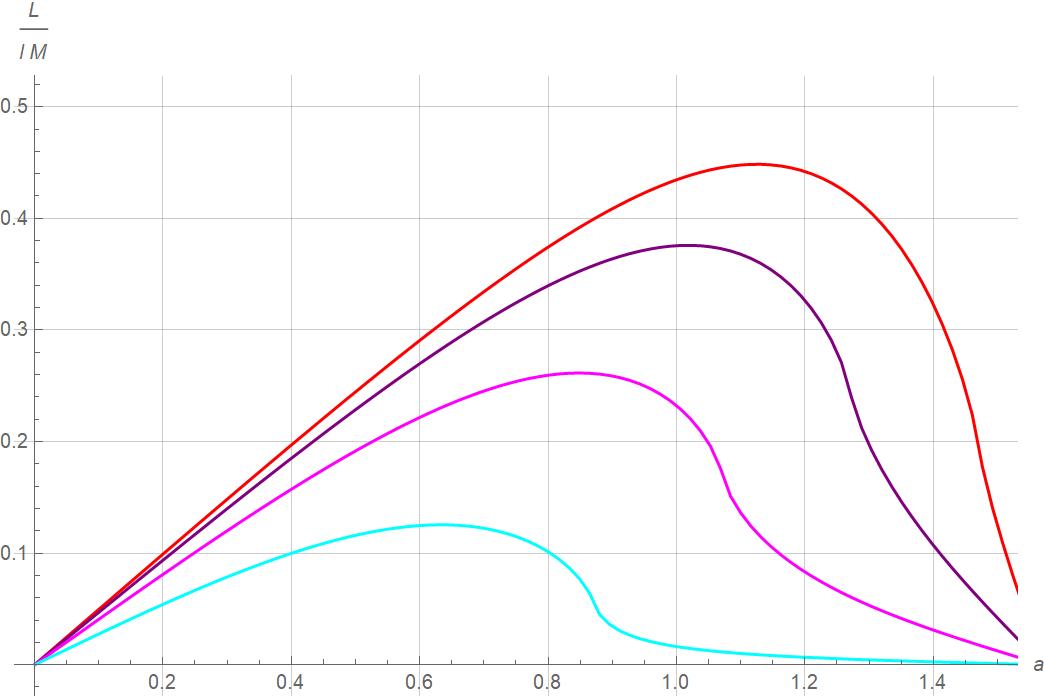}
    \caption{We plot the normalized $L/M l $ as a function of the parameter $a$ and find that it is everywhere smaller than $1$, demonstrating \eqref{higu}. The colors are as in figure \ref{fig: higu}. It is clear that smaller $b$ give higher maxima and for $b=0$ we are back at the GKP-like string.}.
    \label{fig: higu2}
\end{figure}

In Figure \ref{fig: higu} one sees numerically that \eqref{higu} is indeed always satisfied, for each regime. In going to regime 1, which corresponds to the smaller values of $a$ on the left side of \ref{fig: higu}, the induced metric \eqref{larsaninduced} flips sign, as its conformal factor is $r^2 - l^2\sin^2 a$. Indeed at the boundary between the two regimes we get a particle at the speed of light with induced metric that vanishes. This is where the curve hits zero. In Figure \ref{fig: higu2} we plotted the more

 telling normalized relation. As the examples in the next section will make clear, each string is composed of a number $n$ of identical pieces. Each of these pieces has an angular momentum no larger than that of the open straight string with highest angular momentum. This is in line with the `maximal spin' noted in \cite{Noumi2019}.

\subsection{Examples for the three regimes}
Now we can briefly give some examples of what the solutions look like at fixed time in static patch coordinates. For each of the regimes we give an example of a closed string, with $a, b$ determined numerically such that \eqref{closed} is true. In the corresponding plots we also give the numerical results for their energy and angular momentum. Whether these solutions are of sinh- or cosh-Gordon (or Liouville) type depends on the sign of $\cos{4a}-\cos{4b}$, according to the discussion in \cite{larsen1996} translated to our notation. 

\begin{figure}
    \centering
   \begin{subfigure}{0.32\textwidth}
            \centering
            \includegraphics[width=\textwidth]{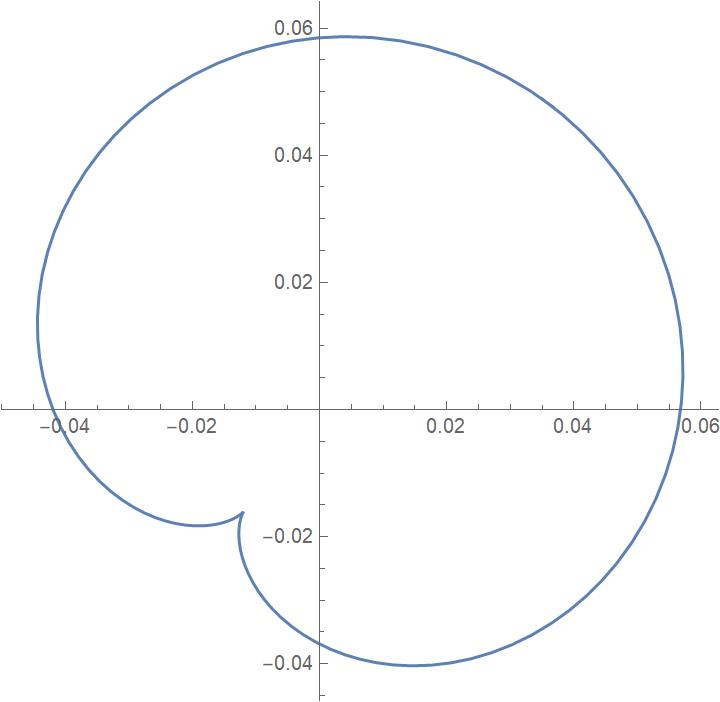}
            \caption[]%
            {{\small Inward cusps}}    
            \label{fig:typ1}
        \end{subfigure}
        \hfill
        \begin{subfigure}{0.32\textwidth}  
            \centering 
            \includegraphics[width=\textwidth]{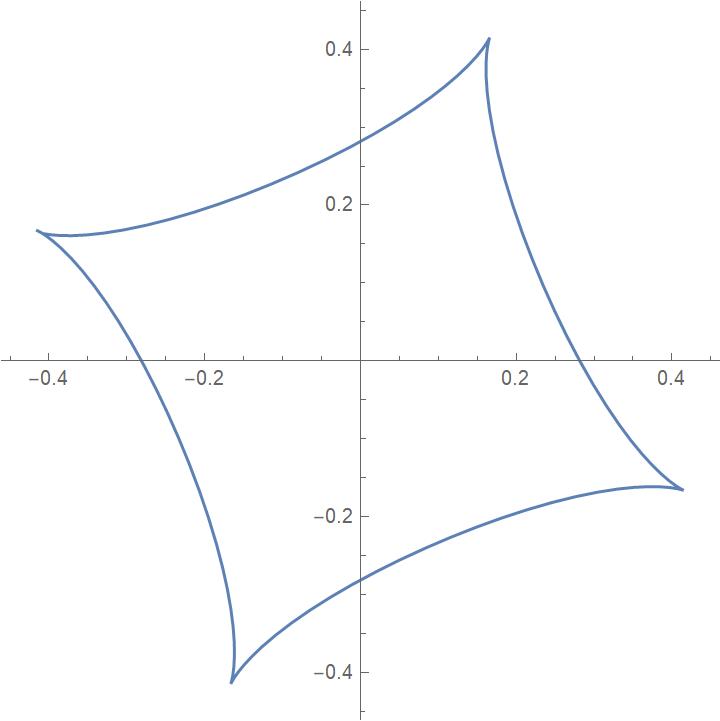}
            \caption[]%
            {{\small Outward cusps}}    
            \label{fig:typ2}
        \end{subfigure}
                \hfill
        \begin{subfigure}{0.32\textwidth}  
            \centering 
            \includegraphics[width=\textwidth]{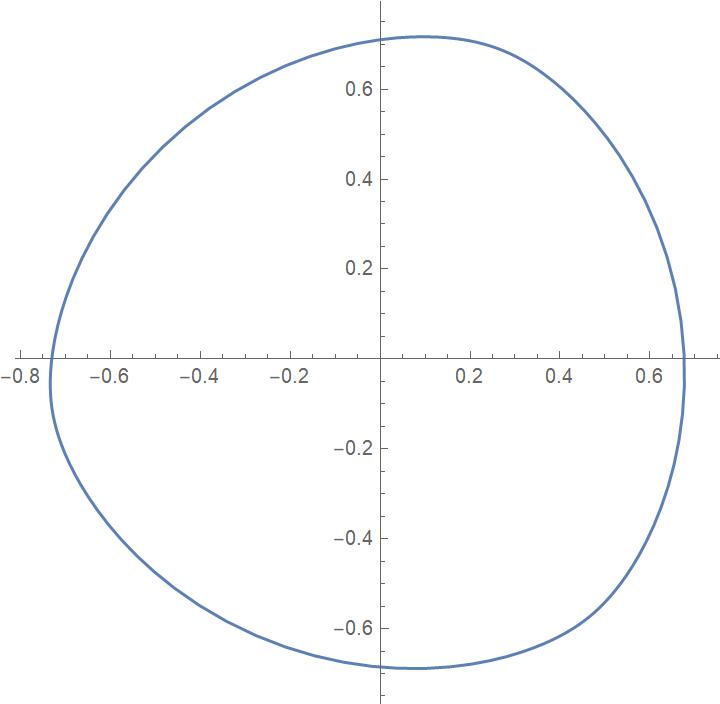}
            \caption[]%
            {{\small No cusps}}    
            \label{fig:typ3}
        \end{subfigure}
    \caption{Examples from each of the three types of solutions in the Larsan-Sanchez class that are shifted by an imaginary half-period. The origin of the plots is that of the static patch and the axes measure the dimensionless coordinate distance $r/l$. }\label{fig:types}
\end{figure}

The first case has $r_a < r_b$ and is characterized by inward cusps. Such solutions are of sinh-Gordon type. A particular example with $n=m=1$ is obtained for $a=0.019997$ and $b=0.060145$ as shown in Figure \ref{fig:typ1}. 

In the second case we have $r_b < r_a < r_c$ and hence there are cusps on the outside. These solutions are of cosh-Gordon type as $\cos4a < \cos 4b$. An example with $m=1$ and $n=4$ is found by taking $a=0.463647$ and $b = 0.256502$ and can be seen in Figure \ref{fig:typ2}. As discussed in \ref{app:Kruc}, these two cases with cusps are the ones that correspond to the Wick rotation of Kruczenski's solutions in $\AdS$ \cite{Kruczenski_2005}.

Finally, there is the case $r_b  < r_c< r_a$, which means there are no cusps. These solutions are again of sinh-Gordon type. An example with $m=1$ and $n=3$ is given by $a=0.852909$ and $b=0.734$. It is shown in Figure \ref{fig:typ3}. Note that the parts of the string inside $r =l/\sqrt{2}$ balance off those parts outside of it, which would otherwise expand toward the horizon. This also helps to understand how the cusps have been smoothed out, which does not occur in $\AdS$. 

\subsection{Expanding and infinite strings}

\begin{figure}
    \centering
    \includegraphics[width = 0.4\textwidth]{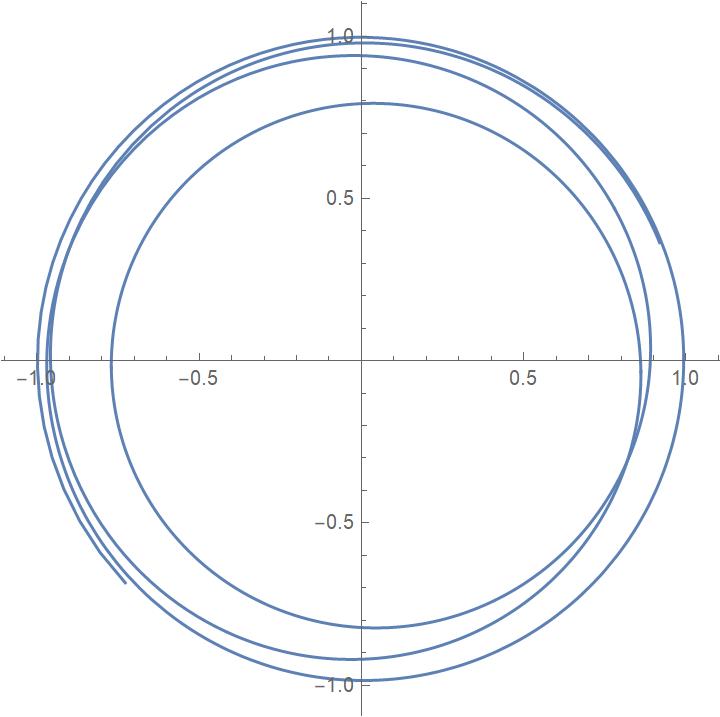}
    \caption{Unshifted solution for $a= \arctan\frac65$ and $b=\frac12 \arcsin2$. The string winds around the horizon as in the case of a stationary string around a Kerr horizon. However, in this case the horizon is static and the string itself rotates uniformly and therefore has infinite mass and is rather unphysical. The string keeps winding around and has only been plotted partly since Mathematica takes more and more time when getting near the horizon at $r/l = 1$.}
    \label{fig:unshifted}
\end{figure}

One type of  Larsen-Sanchez solutions that so far we haven't discussed are the ones where the Weierstrass elliptic functions have not been shifted by the imaginary half-period. Those can describe expanding strings that reach the horizon in finite proper time, such as circular strings that start at $r>l/\sqrt{2}$. The rigidly rotating solutions of this unshifted type moreover extend to the horizon and consequently have infinite mass and angular momentum. Nonetheless, the general behaviour of such a solution near the horizon may still be of relevance. We can see from \eqref{fgeq} that $t'$ diverges as $1/(1-\frac{r^2}{l^2})$. Therefore, to find the shape at fixed $t$, we must let $\tau$ run over a compensating range. This will result in a change 

\begin{equation}
    \Delta \varphi \propto \log \frac{1 + \frac{r}{l}}{1 - \frac{r}{l}}
\end{equation}

which is similar to the logaritmic winding of a static string near the horizon of a Kerr black-hole \cite{Frolov:1988zn, Frolov:1998wf}. In this case however the string itself rotates uniformly and the horizon does not, resulting in unbounded energy and making the solution less physical. One could however imagine a rotating string achieving a certain winding until it possibly reconnects at one of its self-intersections. From the mathematical point of view, one can make a link with classification results of \cite{bychen}, in the specific case of Liouville type of solutions where the Gauss curvature is constant. We discuss this in appendix \ref{app:cheng}. Of course, the energy and angular momentum integrals will diverge, but integrating over finite parts, one can find examples where $Ml < L$. Again this does not contradict \eqref{higu} since these strings do not remain strictly within the static patch, as they stretch ever closer to the horizon. A typical solution of this unshifted kind is given in Figure \ref{fig:unshifted}. 

\section{Nambu-Goto approach}\label{sec:ng}
Instead of starting from the Polyakov action in conformal gauge, in some classical cases it is easier to start instead from the Nambu-Goto action 

\begin{equation}
    S = \int \rmd \tau \rmd \sigma \mathcal{L}_{ng} =  \int \rmd \tau \rmd \sigma \sqrt{(\dot{X}^{\mu} X^{'}_\mu)^2-\dot{X}^\mu\dot{X}_\mu {X^{'}}^{\nu} X^{'}_{\nu}  }
\end{equation}

where contractions are made with the static patch metric \eqref{statpatch}. Here we can take a static gauge $t= \tau$ to find solutions of interest. Simply calculating the equations of motion in the static patch of $\dS_3$ would give three equations that are proportional to each other, due to the reparametrization invariance, so we need to impose an extra condition apart from static gauge. For the rest of this section we will also take $l=1$. 

\subsection{Closed strings near the horizon}\label{sec:ng11}

For closed strings that are single-valued as a function of $\varphi$, in particular they must go around the origin only once, we can fix the invariance by taking a cylindrical static gauge, where $t=\tau$ and $\varphi=\sigma$. The equation of motion 

\begin{equation}
    \partial_\sigma \mathcal{P}^\sigma_t + \partial_\tau \mathcal{P}^\tau_t = 0
\end{equation}

with 

\begin{equation}
    \mathcal{P}^\sigma_t = \frac{\partial \mathcal{L}_{ng}}{\partial t'}, \hspace{0.5 cm} \mathcal{P}^\tau_t = \frac{\partial \mathcal{L}_{ng}}{\partial\dot{t} }
\end{equation}

then becomes, with $l=1$

\begin{equation}\label{ngcyl}\begin{split}
    &\ddot{r}(r^5 - r^3 - r r'^2) + r'' (r(1- \dot{r}^2)+r^5 - 2r^3) + 2r \dot{r}r' \dot{r}'\\&= r^2 - 4r^4 + 5r^6 - 2 r^8  +  \dot{r}^2(4r^4 - r^2)+ r'^2 (2- 6r^2 + 4r^4). \end{split}
\end{equation}

We can use \eqref{ngcyl} to numerically solve for the string motion, given a certain initial shape and velocity. For strings that rotate rigidly, i.e. don't change shape but just increase their value of $\varphi$ linearly in time, we have $r(\tau,\sigma) = r(\sigma- \omega \tau)$. In that case, we end up with a single non-linear ODE for $r$. This ODE determines the shape of rigidly rotating string segment, given a point on that segment and its tangent vector. It shows that the Larsen-Sanchez solutions are in fact the unique rigidly rotating solutions\footnote{Though of course for the ones with cusps the cylindrical parametrization breaks down at the cusps themselves.}. Finding general Larsen-Sanchez solutions in this gauge is more involved, but for a circular string oscillating at frequency $\omega$ we have $r(\tau)$ satisfying

\begin{equation}
       \dot{r} =  (1-r^2)(1-\omega^2 r^2 + \omega^2 r^4)^{\frac12}
\end{equation}

describing the solution in \cite{DeVega1994a}. For $\omega=2$ we have the special static string at $r=1/\sqrt{2}$. For smaller $\omega$ the string moves through the origin and then back to the horizon, whereas for $\omega>2$ the string either oscillates near the origin or it comes in from the near-horizon region, reaches a minimum distance to the origin and expands again towards the horizon. We can now add a perturbation $r = r_0(t) + \varepsilon r_1(\tau, \sigma)$ to this. To first order in $\epsilon$, using that $r_0$ is a circular solution

\begin{equation}\label{ngpert}
    \ddot{r}_1 - r''(1-r^2_0)^2\omega^2 + \frac{8 r^2_0 -2}{r_0}\sqrt{1 - \omega^2 r^2_0(1- r^2_0)}\;\dot{r}_1 - r_1(2+\omega^2 - 10r^2_0 + \omega^2 r^4_0 (3 - 4r^2_0))=0.
\end{equation}

Later on we want to look at an example for which $M l > L$, which means we will have to take a solution that is not close to the circular one. Still, as we noticed previously it means that the string will have to move near the horizon, so it is useful to look at what happens in that regime, even in the more tractable circular case. To understand perturbations in this limit, we can take $r\to 1$ in \eqref{ngpert} and find

\begin{equation}
    \ddot{r}_1 + 6 \dot{r}_1 + 8 r_1 = 0
\end{equation}

showing that irrespective of the precise shape of the perturbations, to a static patch observer they will appear exponentially damped as $r_1(t) \sim e^{\lambda t}$ with $\lambda = -2$ or $-4$. This is due to the fact that the static patch time diverges exponentially as a function of the global time, in the neighborhood of the horizon. The above describes the perturbation in the coordinate distance $r$. Near the horizon however, for the circular string, $t$ itself behaves as $t \approx -\frac12 \log(1-r)$, so the physical distance $\delta s$ evolves as

\begin{equation}
    \delta s = \frac{\delta r}{(1-r^2)^{1/2}} \sim \rme^{(\lambda + 1)t}  
\end{equation}

which stills decreases exponentially. For more general perturbations, the above gets involved and perhaps a more technical approach along the lines of \cite{Frolov:1996be} could be beneficial. 

To find an example with $M > L$, we need to move away from the non-rotating circular case and take as initial position $r$ and velocity $\dot{r}$ something rapidly moving\footnote{While still having a Lorentzian worldsheet.} and necessarily close to the horizon, like

\begin{equation}\label{numex}
    r(0, \sigma) = 0.012 \sin8\sigma + 0.98, \hspace{0.5cm} \dot{r}(0,\sigma) = 0.03 \cos 8 \sigma.
\end{equation}

The energy and angular momentum are determined by

\begin{equation}
    M = \int \rmd \sigma \mathcal{P}^\tau_t , \hspace{0.5cm} L = \int \rmd \sigma \mathcal{P}^\tau_\varphi
\end{equation}

with

\begin{equation}\label{ngconserved}
    \mathcal{P}^\tau_t = \frac{r^2- r^4+ r'^2}{\sqrt{r'^2 + r^2(1- \frac{\dot{r}^2}{1-r^2} ) -r^4}}, \hspace{0.5cm} \mathcal{P}^\tau_\varphi = \frac{r^2 \dot{r} r'}{(1-r^2)\sqrt{r'^2 + r^2(1- \frac{\dot{r}^2}{1-r^2})-r^4}}.
\end{equation}

\begin{figure}
    \centering
   \begin{subfigure}{0.4\textwidth}
            \centering
            \includegraphics[width=\textwidth]{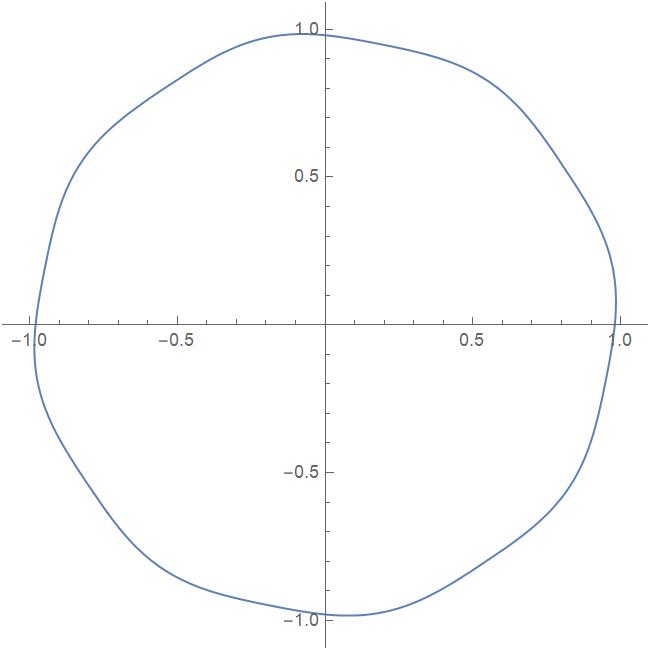}
            \caption[]%
            {{\small string at $\tau = 0$}}    
            \label{fig:nginit}
        \end{subfigure}
        \hspace{0.5cm}
        \begin{subfigure}{0.4\textwidth}  
            \centering 
            \includegraphics[width=\textwidth]{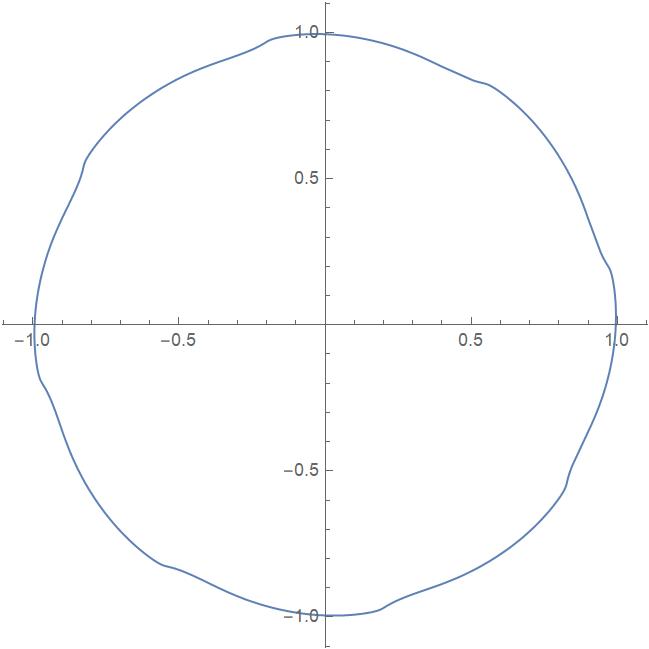}
            \caption[]%
            {{\small string at $\tau = 0.7$}}    
            \label{fig:ngfin}
        \end{subfigure}
        \vskip\baselineskip
        \begin{subfigure}{0.4\textwidth}   
            \centering 
            \includegraphics[width=\textwidth]{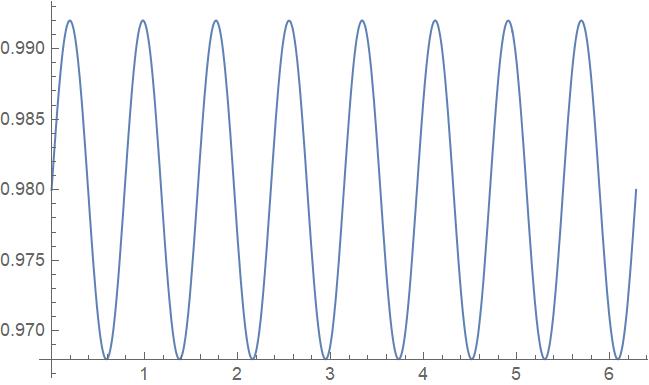}
            \caption[]%
            {{\small radius at $\tau=0$}}    
            \label{fig:nginitzoom}
        \end{subfigure}
        \hspace{0.5cm}
        \begin{subfigure}{0.4\textwidth}   
            \centering 
            \includegraphics[width=\textwidth]{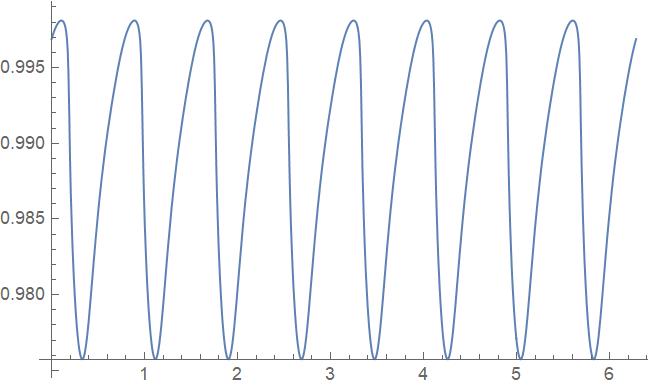}
            \caption[]%
            {{\small  radius as at $\tau=1.4$}}    
            \label{fig:ngfinzoom}
        \end{subfigure}
    \caption{Numerical example with $M l > L$, showing that this bound need not hold for strings that expand indefinitely towards the horizon. Initial conditions are $r(0, \sigma) = 0.012 \sin8\sigma + 0.98$ and $\dot{r}(0,\sigma) = 0.03 \cos 8 \sigma$. The axes give the dimensionless coordinate distance $r/l$, with horizon at $r/l=1$. One can see how the string moves towards the horizon and how in doing so the outer extrema get slowed down most, while the inner parts continue to move.}\label{fig:ngnumeric}
\end{figure}

The square roots are real whenever the worldsheet is indeed Lorentzian. For \eqref{numex}, we find numerically $M= 1.89$ and $L= 1.62$. So this goes to show that, for strings with center of mass at the origin, the bound $M > L$ can be violated when the string will end up in the horizon. For $\tau$ not too large, we can solve \eqref{ngcyl} numerically and the shape changes as in Figure \ref{fig:ngnumeric}.

The string as a whole moves outwards and the amplitude of the perturbation grows smaller in the coordinate $r$. We see also that the tips of the perturbation get stuck on the horizon, while the inner parts of the string continue to move by, resulting in a change in shape. By taking

\begin{equation}\label{numex2}
    r(0, \sigma) = \frac{1}{n} \sin n\sigma + (1-\frac{2}{n}), \hspace{0.5cm} \dot{r}(0,\sigma) = \frac{1}{\sqrt{n}} \cos n \sigma.
\end{equation}

with $n\in \mathbb{N}$ large, we can make $L/M $ arbitrarily large, growing like $\sqrt{n}$, while still having a Lorentzian worldsheet.

Finally, from \eqref{ngconserved} one can also see that any string, with $L\geq M $ will need to have a part outside of $1/\sqrt{2}$. Previously, in the discussion around \eqref{larsanpos}, we only obtained this for rigid strings. Here we see it generally, noting that having a Lorentzian worldsheet imposes

\begin{equation}
    \frac{r^2 \dot{r}^2}{1-r^2} \leq r^2 - r^4 + r'^2.
\end{equation}

This means that even locally $\mathcal{P}^\tau_t > \mathcal{P}^\tau_\varphi $ unless $r' > \dot{r}$. If that is the case, then 

\begin{equation}
    \frac{\mathcal{P}^\tau_\varphi}{\mathcal{P}^\tau_t} < \frac{r^2r'^2}{(1-r^2)(r^2 - r^4 + r'^2)},  
\end{equation}

where the RHS will be less than $1$ if $r < 1/\sqrt{2}$.

\subsection{Open strings with blobs of mass}

One could also wonder about the movement of open strings. In this case it is natural to take $t=\tau$ and impose the orthogonality condition of vanishing $\dS$ inner product $g_{\mu\nu} \dot{X}^\mu X'^\nu=0$. Then we obtain as equation of motion 

\begin{equation}
 \partial_\tau \mathcal{P}^\tau_t =0, \hspace{0.5cm} \mathcal{P}^\tau_t = (1-r^2)\frac{X'^2}{\sqrt{-\dot{X}^2X'^2}} .
\end{equation}

with $X$ denoting the coordinates and taking as inner product the $\dS$ one. Solutions are found by considering some $\mathcal{P}^\tau_t = K(\sigma)$. Near the ends of the open string, where $\dot{X}^2=0$, due to the vanishing of $\mathcal{P}^\sigma_\varphi$ and $\mathcal{P}^\sigma_r$, we also necessarily have $ X'^2=0$ whenever $K$ remains finite. In this way one can also numerically evolve strings with a given initial shape $r(\sigma)$ and velocity profile determined by $K(\sigma)$ by solving both the above and the orthogonality constraint. This simple-minded approach seems to run in to serious numerical errors rather quickly. Nevertheless, at least at small times one can consider a straight string that has been given a higher angular velocity for its inner parts than in the rigid rotor. The string starts to bulge and the endpoints move outward towards the horizon.

One can however see that the only rigidly rotating solution through the origin is the GKP-like rigid rotor. This can be done by forgetting about the orthogonality constraint and instead parametrizing as $r(\sigma)$ and $\varphi = \omega \tau + \varphi(\sigma)$. The equation of motion can be rewritten as an equation for $\varphi(r)$ alone, 

\begin{equation}\label{openphir}\begin{split}
    &(-2 + (6 + \omega^2)r^2-4(1+\omega^2)r^4)\frac{\rmd \varphi}{\rmd r} + r^2(1-r^2)^2(-1+2r^2)(\frac{\rmd \varphi}{\rmd r})^3\\&=r(1-r^2)(1-(1+\omega^2)r^2)\frac{\rmd^2 \varphi}{\rmd r^2}.
\end{split}
\end{equation}

The straight rotor is obtained by $\varphi'(r)=0$ everywhere. Apart from this solution, \eqref{openphir} has no other solutions that can be Taylor-expanded around the origin. By uniqueness, the other solutions that do not reach the origin again describe parts of the Larsen-Sanchez strings.

To change the setup and gain some more intuition one could also imagine modeling a rotating string as having an inner part consisting of a straight, rigidly rotating open string and outer parts that are described by some blobs of mass. In the extremely simplified case, we would be adding a point-mass to each end of the open GKP-like string string rotating at angular velocity $\omega$. This configuration, when having the masses $M$ at radius $r_0$ has total energy (still setting $l=1$).

\begin{equation}
    E = 2 M \frac{1-r^2_0}{\sqrt{1-(1+\omega^2)r^2_0}} + \int^{r_0}_{-r_0} \rmd r \frac{\sqrt{1-r^2}}{1-(1+\omega^2) r^2}.
\end{equation}

The total angular momentum is

\begin{equation}
    L = \frac{2Mr^2_0 \omega}{\sqrt{1-(1+\omega^2)r^2_0}} + \int^{r_0}_{-r_0} \rmd r \frac{r^2 \omega}{\sqrt{1-r^2}\sqrt{1-(1+\omega^2) r^2}}.
\end{equation}

In equilibrium one needs to solve the equation of motion for $r$,  $\partial_r \mathcal{L}=0$, for the equilibrium length $r_0$. This tells us that the tension balances the centripetal acceleration

\begin{equation}
    \frac{2 M r (1 + \omega^2)}{\sqrt{1 + (1 + \omega^2) r^2}} - 2 \sqrt{\frac{1 - (1+ \omega^2) r^2}{1-r^2}}=0.
\end{equation}

\begin{figure}
    \centering
    \includegraphics[width = 0.6\textwidth]{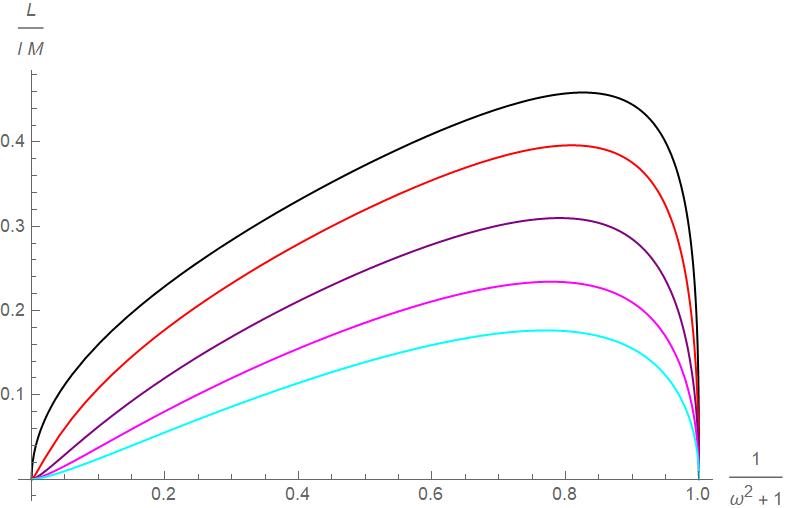}
    \caption{In black $L/M$ for the straight open string that rigidly rotates, as in \cite{Noumi2019}. The other curves are for increasing masses added at the two ends of the string, ranging from 0.25 in red to 1 in cyan.}
    \label{fig: blob}
\end{figure}

We solve for $r_0$ and plug into $E$ and $L$. The result is that the ratio $L/E$ is always smaller than that of the straight rotor with same angular velocity as seen in Figure \ref{fig: blob}. If we imagine modeling the outer part of the string by some blob of mass attached to an inner straight part, it means that for high $L/E$ the configuration can only be stabilized by letting the blob expand towards the horizon, in line with what we argued previously. For the straight string itself, the $L/E$ ratio is bounded by $0.4588$ at $\omega = 0.4588$, where $E-L/\omega =0$. Both $E$ and $L$ reach a maximum at this point, as previously reported in \cite{Noumi2019}. A similar conclusion, in the context of $\dS$ field theory, was reached for spinning mesons in \cite{Chu:2016pea}.

\section{Segmented strings}\label{sec:Gubser}
In \cite{Vegh:2015ska} and \cite{Callebaut:2015fsa} it was found that the evolution of classical strings in $\AdS_3$ could be approximated by exact solutions given in terms of a finite number of kinks moving at the speed of light, pairwise lying in a totally geodesic $\AdS_2$ subspace and colliding with each other after some time, thereby changing the shape of the string. Soon after, it was realized by Gubser in \cite{Gubser:2016wno} that such solutions could be completely described by an algebraic evolution law for discrete lightcone coordinates on the worldsheet. Since this is easily implemented on a computer, we intend to follow this procedure here for $\dS_3$. 

We will begin with reviewing the evolution law of \cite{Gubser:2016wno}. In the next section, we build further on this work, by obtaining a simple expression for the conserved energy and angular momentum of the string. For given initial configurations with 4, 6 and 8 collisions we generate a random sample of initial velocities. The generic movement is chaotic and the energy rather than angular moments in general determines whether the string will exit the static patch, after which the evolution law will break down. Finally, inspired by section \ref{sec:ng11}, we give an explicit example of a segmented string with its angular momentum larger than its mass. The discrete approach makes it easy to see how the string expands as it rotates and leaves the static patch, until at some point the kinks become causally disconnected and do not collide again.

\subsection{Algebraic evolution in flat space and \texorpdfstring{$dS_3$}{TEXT}}

Here we briefly review the ideas leading to the Gubser evolution law \cite{Gubser:2016wno}. Let us begin with a string in flat space, where we can decouple the evolution in terms of left- and right-movers. A worldsheet of the form

\begin{equation}
    X(\tau,\sigma) = Y_L(\tau+\sigma)+Y_R(\tau-\sigma)
\end{equation}

with $Y_L$ and $Y_R$ lightlike trajectories will solve the Polyakov equations of motion with constraints. Some neat examples are when these trajectories move around circles of different radii. In that case the resulting curve is an epicycloid or hypocycloid depending on the relative direction of motion of $Y_L$ and $Y_R$. The strings rigidly rotate and are the flat space analogs of the Larsen-Sanchez strings in regime 1 and 2 respectively, see section \ref{sec: larsan}. Now one can discretize these lightlike trajectories, by making them piecewise linear and the resulting string still solves the equations of motion. It can be given in terms of discrete lightlike coordinates $(i,j)$ on the worldsheet

\begin{equation}
    X_{i,j} = Y_{L,i}+Y_{R,j}.
\end{equation}

\begin{figure}
    \centering
   \begin{subfigure}[b]{0.3\textwidth}
            \centering
            \includegraphics[width=\textwidth]{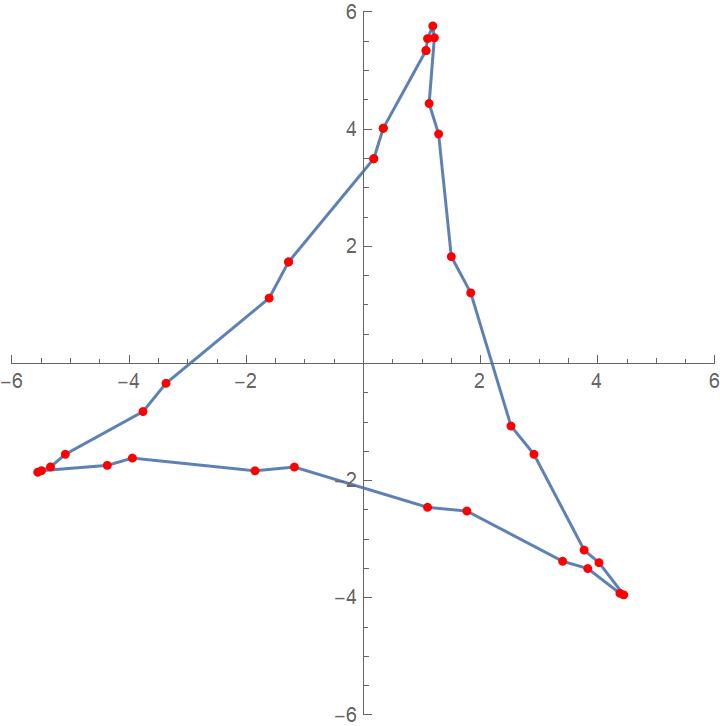}
            \caption[]%
            {{\small before collision}}    
        \end{subfigure}
        \hspace{0.2cm}
        \begin{subfigure}[b]{0.3\textwidth}  
            \centering 
            \includegraphics[width=\textwidth]{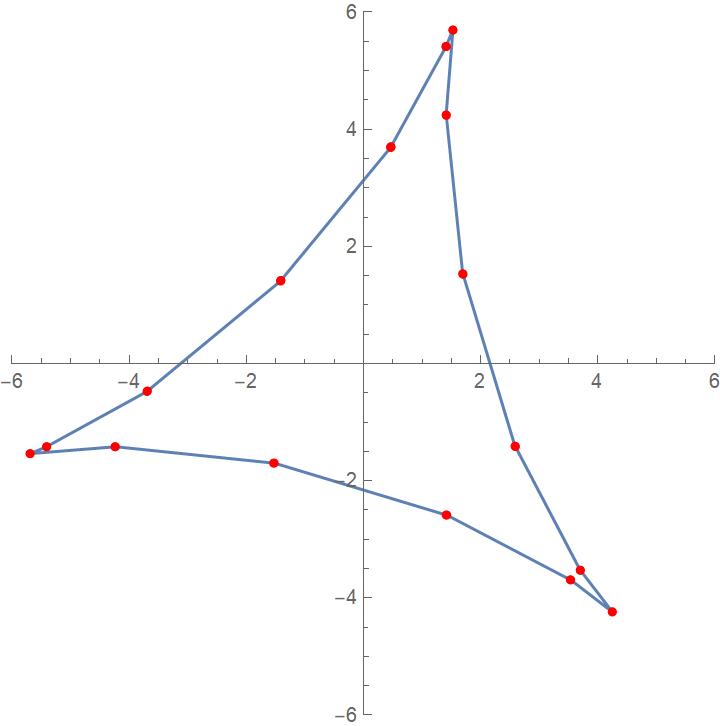}
            \caption[]%
            {{\small kink collision}}    
        \end{subfigure}
        \hspace{0.2cm}
        \begin{subfigure}[b]{0.3\textwidth}   
            \centering 
            \includegraphics[width=\textwidth]{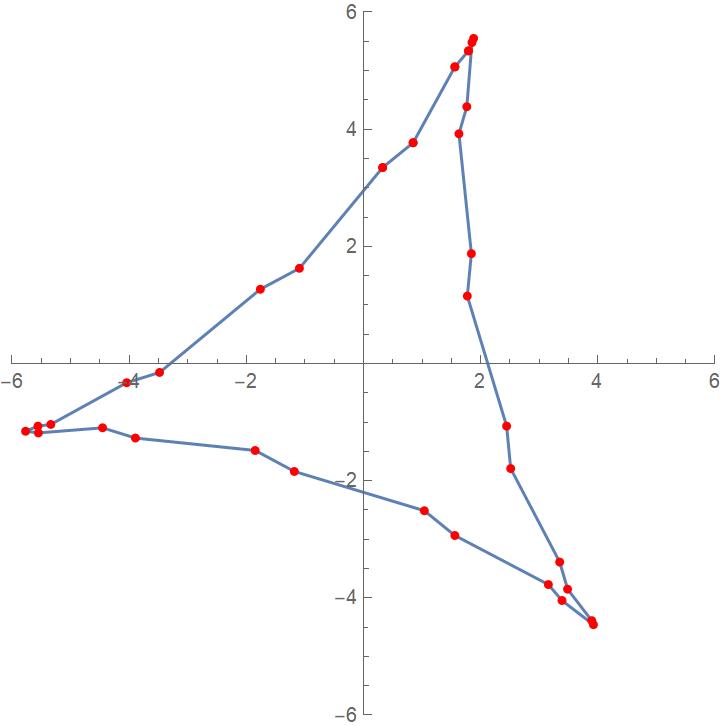}
            \caption[]%
            {{\small after collision}}    
        \end{subfigure}
    \caption{This segmented string in flat space is obtained by taking $Y_L$ of radius 4 and $Y_R$ or radius 3. There are 32 kinks and 16 collisions, which in this special case occur all at the same time. This string is an exact solution and approximates a rigidly rotating hypocycloid, which is the flat space analog of the regime 2 Larsen-Sanchez solutions of section \ref{sec: larsan}. }\label{fig:flatseg}
\end{figure}

These events $X_{i,j}$ are the kink collisions at which two kinks collide and subsequently move apart. The entire string at fixed time is then the collection of these ordered kinks, connected by straight line segments. An example is given in Figure \ref{fig:flatseg}. For closed strings there is a periodicity $X_{i+n,j-n}=X_{i,j}$ for some $n$.

A very analogous approach can be now be considered in $\AdS_3$ and $\dS_3$ \cite{Vegh:2015ska, Callebaut:2015fsa, Gubser:2016wno}. In this case, we still have kinks moving at the speed of light, in straight line segments in embedding space. A collision happens at points $X_{i,j}$, after which one kink moves by a lightlike amount $\Delta_{L,i,j}$ and similarly the other moves by $\Delta_{R,i,j}$. As opposed to flat space, these $\Delta$ now generally depend on both $i$ and $j$. The three quantities $X$, $\Delta_L$ and $\Delta_R$ determine a plane in embedding space and, upon taking the intersection with the hyperboloid, an $\AdS_2$ or $\dS_2$ subspace in which the two adjacent kinks move. The straight lines connecting the kinks in flat space are therefore replaced by fixed time sections of totally geodesic $\AdS_2$ and $\dS_2$ subspaces. Given the causal wedge

\begin{equation}
    X_{i,j}, \hspace{0.5cm} X_{i+1,j} = X_{i,j}+\Delta_{L,i,j}, \hspace{0.5cm} X_{i,j+1} =X_{i,j}+\Delta_{R,i,j}
\end{equation}

Gubser's evolution equation, stated here for $\dS_3$, still taking the $\dS$ length $l=1$, uniquely determines $X_{i+1,j+1}$ as

\begin{equation}\label{gubser}
    X_{i+1,j+1} = X_{i,j} +2 \;\frac{\Delta_L + \Delta_R - (\Delta_L\cdot \Delta_R)X_{i,j}}{2+  \Delta_L\cdot \Delta_R}
\end{equation}

where the inner product is that of the embedding space $\mathbb{R}^{1,3}$ and we omitted the indices on $\Delta_{L,R}$ to avoid clutter. It was already noted in \cite{Gubser:2016wno} that if $\Delta_L\cdot \Delta_R < -2$ the kinks are causally disconnected, which is possible due to the expansion of $\dS_3$, and there will not be a future collision. The evolution \eqref{gubser} itself is not to be used there, since it would give a collision at an earlier time.

\subsection{Mass and angular momentum}

Having reviewed the result of \cite{Gubser:2016wno}, we will now continue by obtaining expressions for the energy and angular momentum of segmented strings in $\dS_3$. We will first take a step back from the discrete coordinates to continuous ones, and find the conserved quantities in the usual approach. The result will be entirely in terms of the discrete coordinates and is easily checked to be invariant under \eqref{gubser}. 

First we start by extending the evolution equation to describe the $\dS_2$ subspace in which the kinks move. This is easily done by taking

\begin{equation}\label{gubs2}
    X(\alpha, \beta) = X_{i,j} +2 \;\frac{\alpha\Delta_L + \beta \Delta_R - \alpha \beta(\Delta_L\cdot \Delta_R)X_{i,j}}{2+ \alpha\beta \Delta_L\cdot \Delta_R}.
\end{equation}

Clearly $X(\alpha,0)$ and $X(0,\beta)$ are the kink trajectories emerging from the collision $X_{i,j}$. Energy corresponds to the symmetry $\delta X^0 = X^1$ and $\delta X^1 = X^0$, whereas angular momentum corresponds to $\delta X^2 = -X^3$ and $\delta X^3 = X^2$  The conserved currents obtained from the Nambu-Goto action are messy, but along the piecewise linear contour consisting of the line segments $\alpha=0$ and $\beta=0$, they simplify, since

\begin{equation}
    \partial_\alpha X |_{\beta = 0} = \Delta_L , \hspace{0.5cm} \partial_\beta X |_{\alpha = 0} = \Delta_R.
\end{equation}

The square root of the induced metric becomes $|\Delta_L \cdot \Delta_R|$ there. Using moreover that the $\Delta$ are null and therefore $\text{sg}(\Delta_L \cdot \Delta_R) = -1$,  we find that the contribution of either of these two line segments to the energy is simply

\begin{equation}\begin{split}
    M &= \frac12 \int^1_0 \rmd \xi (X^0\partial_\xi X^1- X^1 \partial_\xi X^0) \\
        &= \frac{1}{2}( X^1_{i,j}+ \Delta^1_{i,j})X^0_{i,j} - (X^0_{i,j}+ \Delta^0_{i,j})X^1_{i,j}
\end{split}
\end{equation}

with $\xi = \alpha$  $(\beta)$ whenever $\beta$  $(\alpha)$ is zero and correspondingly in the last line we have $\Delta = \Delta_L$ or $\Delta_R$ respectively. Similarly for the angular momentum,

\begin{equation}\begin{split}
    L &= \frac12 \int^1_0 \rmd \xi (X^2\partial_\xi X^3- X^3 \partial_\xi X^2)\\
    &= \frac{1}{2}( X^2_{i,j}+ \Delta^2_{i,j})X^3_{i,j} - (X^3_{i,j}+ \Delta^3_{i,j})X^2_{i,j}.\end{split}
\end{equation}

The total conserved quantities for the segmented string in $\dS_3$ are then found by taking any closed contour $\gamma$ on the (discretized) worldsheet, consisting of such $\alpha$ and $\beta$ pieces and summing their contributions given above.  For instance, in the next section we will evolve strings from initial conditions $X_{i, -i}$ and $X_{i+1,-i}$, with $i$ periodic modulo some integer $n$. In that case the total mass and angular momentum are obtained from the initial conditions as 

\begin{equation}\label{segmass}
    M = \frac12 \sum^{n}_{k=1} ( (X^1_{k+1,-k}+X^1_{k,-k+1})X^0_{k,-k} - (X^0_{k+1,-k}+X^0_{k,-k+1})X^1_{k,-k})
\end{equation}

\begin{equation}\label{segang}
    L = \frac12 \sum^{n}_{k=1} ( (X^2_{k+1,-k}+X^2_{k,-k+1})X^3_{k,-k} - (X^3_{k+1,-k}+X^3_{k,-k+1})X^2_{k,-k})
\end{equation}

We made brief use of the continuous Nambu-Goto description only to arrive at the right expressions. These expressions however clearly make sense within the discrete description alone and can be checked to be indeed invariant under \eqref{gubser}.

\subsection{Randomly generated strings}

Since they are easily treated in a numerically precise way, we can generate some random segmented strings and look at their energy and angular momentum, comparing it with how long they are expected to stay within the static patch. To do this, we take a simple approach. For 4, 6 and 8 specified initial collisions $X_{i,-i}$ at embedding space time $t=0$ we will randomly generate future collisions $X_{i+1,-i}$. Through \eqref{segmass} and \eqref{segang} we obtain the conserved quantities.

We take the initial conditions to be point-symmetric around the origin of the $X^2X^3$-plane. This is a quick way to ensure that the string has center of mass at the origin. For the $X_{i,i}$ we choose a configuration that is close enough to the origin, so that it does not immediately escape the static patch, yet far enough from it so as to not reduce its dynamics to that of a string in flat space. We also did not take completely symmetric conditions, such as a square, which would result in periodic motion as noticed by \cite{Callebaut:2015fsa} and zero angular momentum. 

\begin{figure}
    \centering
   \begin{subfigure}{0.26\textwidth}
            \centering
            \includegraphics[width=\textwidth]{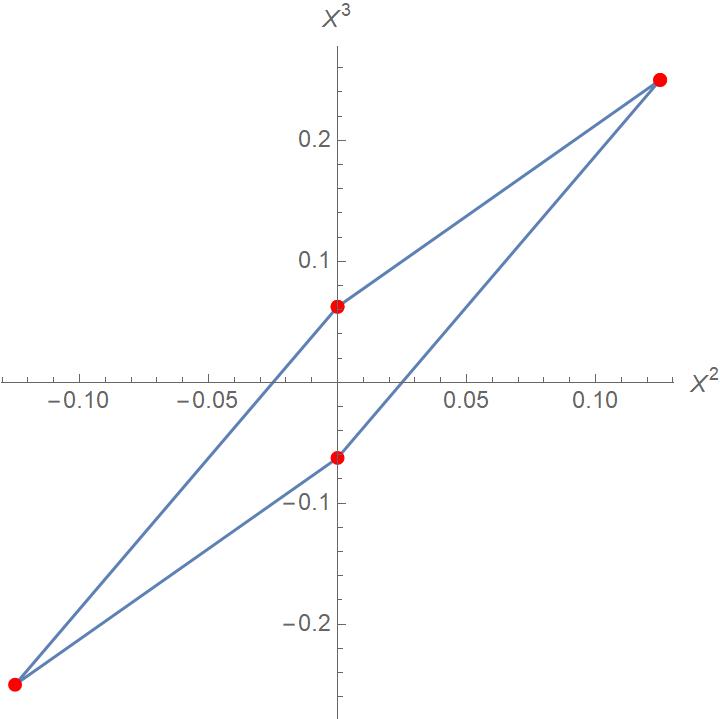}
            \caption[]%
            {{\small 4 initial collisions}}    
            \label{fig:in1}
        \end{subfigure}
        \hspace{0.9cm}
        \begin{subfigure}{0.26\textwidth}  
            \centering 
            \includegraphics[width=\textwidth]{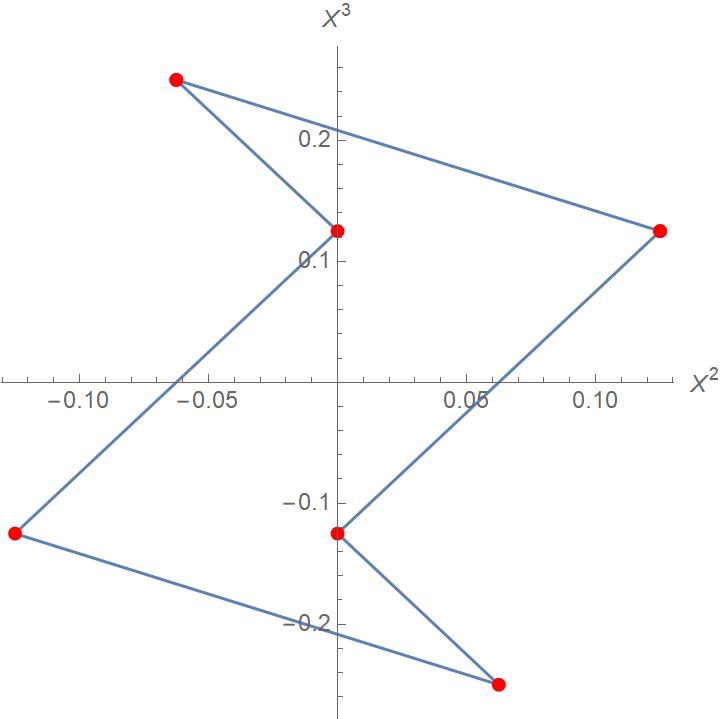}
            \caption[]%
            {{\small 6 initial collisions}}    
            \label{fig:in2}
        \end{subfigure}\hspace{0.9cm}
        \begin{subfigure}{0.26\textwidth}   
            \centering 
            \includegraphics[width=\textwidth]{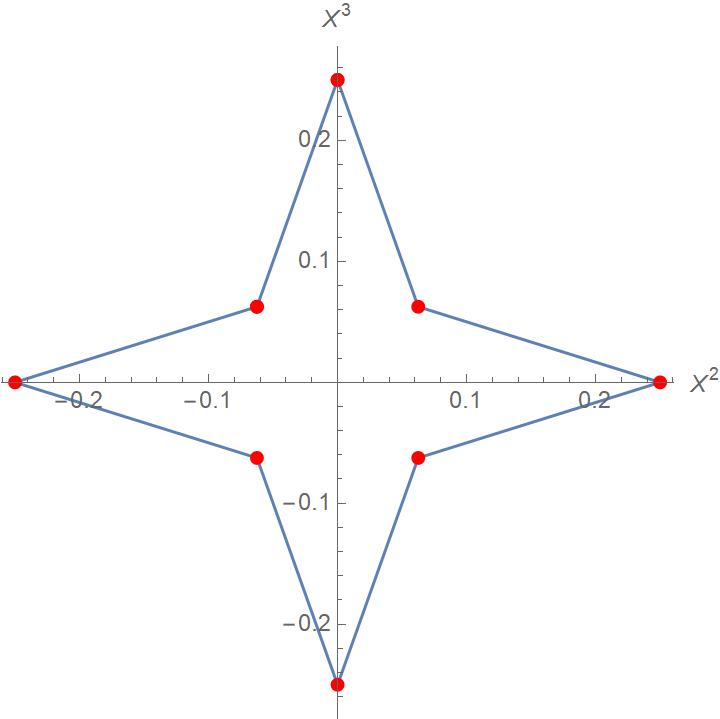}
            \caption[]%
            {{\small 8 initial collisions}}    
            \label{fig:in3}
        \end{subfigure}
        \vskip\baselineskip
        \begin{subfigure}{0.3\textwidth}   
            \centering 
            \includegraphics[width=\textwidth]{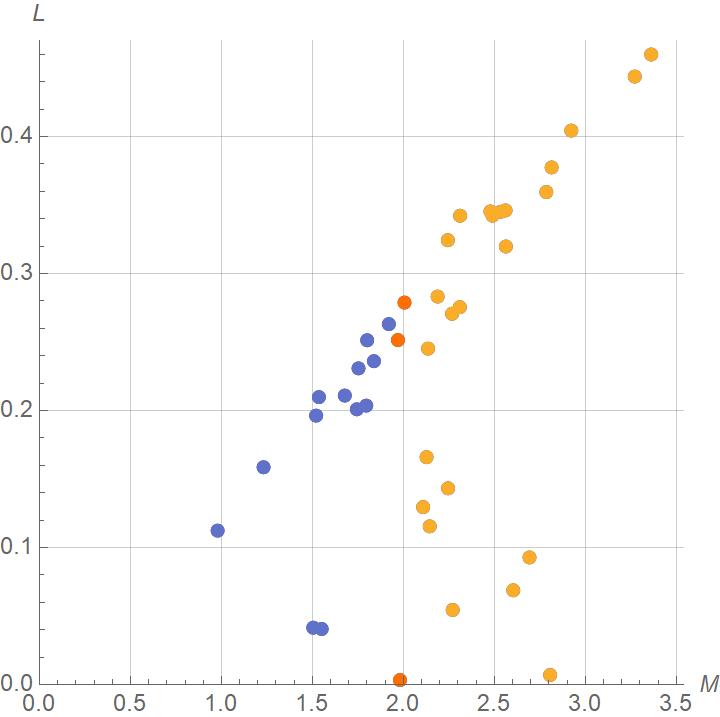}
            \label{fig:scat1}
        \end{subfigure}
        \hspace{0.1cm}
        \begin{subfigure}{0.3\textwidth}   
            \centering 
            \includegraphics[width=\textwidth]{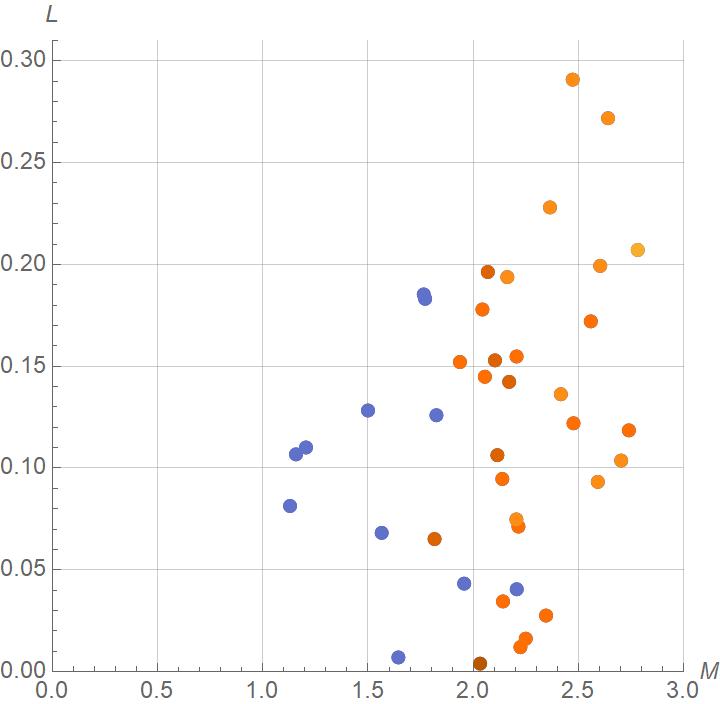}
            \label{fig:scat2}
        \end{subfigure}\hspace{0.1cm}
        \begin{subfigure}{0.34\textwidth}   
            \centering 
            \includegraphics[width=\textwidth]{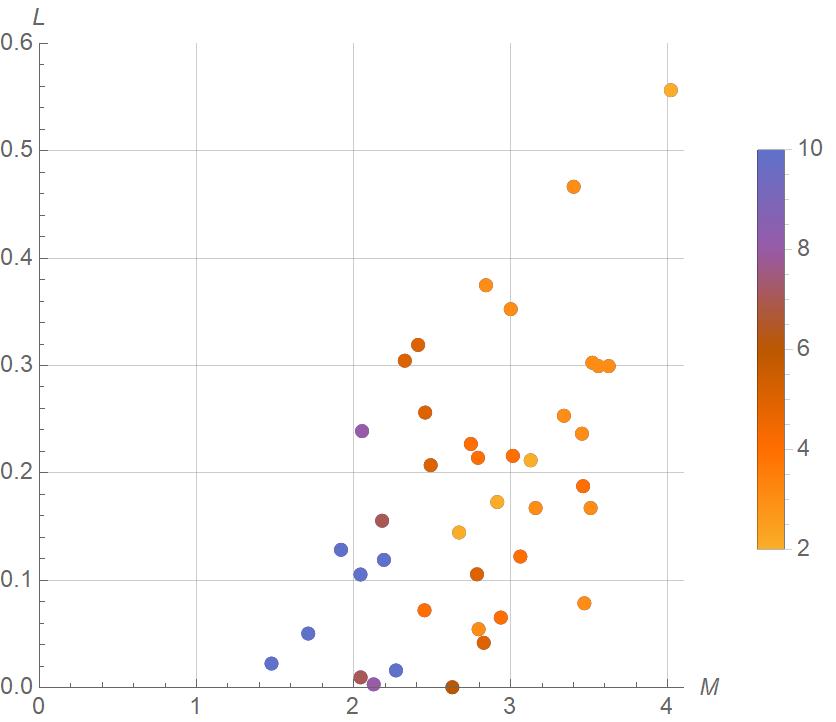}
            \label{fig:scat3}
        \end{subfigure}
    \caption{Considering strings with the above initial collision data $X_{i,-i}$, we generate random $X_{i+1,-i}$, lightlike separated from the initial ones and lying on the $dS_3$ hyperbola. This then fully determines the motion of the string. For each of the initial configurations we show the angular momentum $L$ versus energy $M$ for a sample of 40 thus randomly generated segmented strings. The color scale shows for how many iterations the string remains within the static patch. We stopped after 10 iterations, so that if the string remains within $r\leq 1$ for longer than that, it is also colored in blue. }\label{fig:scatters}
\end{figure}

The initial configurations are shown, for 4 initial collisions $X_{i,-i}$ in Figure \ref{fig:in1}, 6 collisions in Figure \ref{fig:in2} and 8 in Figure \ref{fig:in3}. Different randomly generated $X_{i+1,-i}$ result in strings with different conserved quantities. We take a random sample of size 40 for each case and display energy $M$ versus angular momentum $L$ in the scatterplots of Figure \ref{fig:scatters}. The colors of the points indicate how long the segmented string remains within the static patch. We performed ten iterations of the evolution equation so that the color is determined by an integer between 1 and 10. Strings in blue are the ones that remain inside the static patch for at least ten iterations and are likely to be stable. 

It is notable how the colors are clearly most sensitive to the total energy $M$, as expected. Highly energetic strings, even with low angular momentum are likely to escape to the horizon. All of these randomly generated strings have a rather low angular momentum to energy ratio, but those with larger $L/M$ mostly have both large $M$ and $L$, in line with the expectation that they will leave the static patch.

\subsection{An explicit example with \texorpdfstring{$M l < L$}{TEXT}}\label{sec: segex}

In this section we again wish to find, as in section \ref{sec:ng11}, an explicit example of a string that has angular momentum higher than its energy, violating the bound \eqref{higu}. Previously the numerical evolution of the solution became erratic rather quickly. The method of segmented strings should allow us to understand more exactly its behavior. We will start with a configuration in embedding space at time $t=0$, point-symmetric around the origin of the $X^2X^3$-plane. As before, we want a string that is close to the static patch horizon and has many wiggles. We therefore consider an initial configuration of 32 kink collisions, determined by

\begin{equation}\label{segin1}
    X_{i, -i} = \begin{cases}(0, \sqrt{0.19}, 0.9 \cos{\frac{\pi i}{16}}, 0.9 \sin{\frac{\pi i}{16}}),\quad& i \text{ even} \mod{32}\\
    (0, \sqrt{0.51}, 0.7 \cos{\frac{\pi i}{16}}, 0.7 \sin{\frac{\pi i}{16}}),\quad& i \text{ odd } \mod{32}.
    \end{cases}
\end{equation}

\begin{figure}[ht]
    \centering
   \begin{subfigure}{0.4\textwidth}
            \centering
            \includegraphics[width=\textwidth]{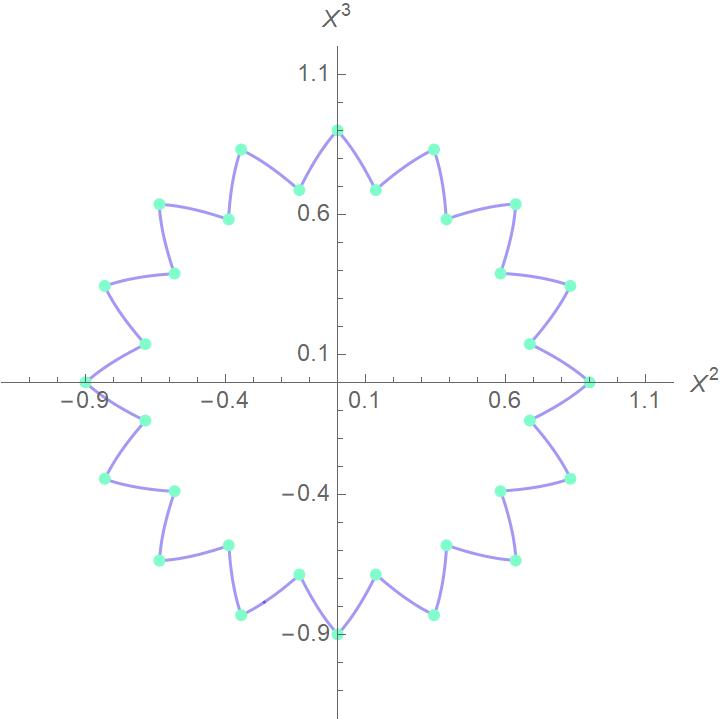}
            \caption[]%
            {{\small Initial configuration at embedding space time $t=0$ with the 32 dots indicating the kink collisions}}    
            \label{fig:exit1}
        \end{subfigure}
        \hspace{0.5cm}
        \begin{subfigure}{0.4\textwidth}  
            \centering 
            \includegraphics[width=\textwidth]{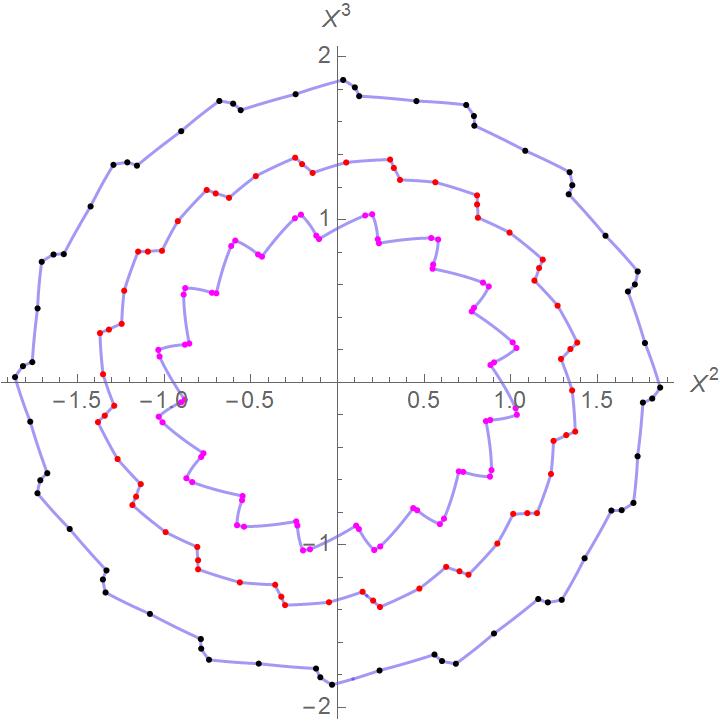}
            \caption[]%
            {{\small Configurations at embedding space times $t=0.6$, $1.2$ and $1.8$. There are 64 kinks at each time}}    
            \label{fig:exit2}
        \end{subfigure}
    \caption{The segmented string with initial configuration determined by \eqref{segin1} and \eqref{segin2} has angular momentum larger than its energy and expands towards the horizon. After 3 iterations the string has partially left the static patch. After 10 iterations the different collisions are causally disconnected and the evolution law breaks down. On the right we see the configuration at intermediate times, when two kinks have come out of each collision and move on to collide with the next neighbor.}\label{fig:exit}
\end{figure}

For the first step we now want to generate 32 new collisions. To this end, we consider $X_{0,0}$ and $X_{1,-1}$. We need to find a suitable new collision that lies on the $\dS_3$ hyperbola and lies on a null ray coming from each of these previous two collisions. We do the same thing for $X_{0,0}$ and $X_{-1,1}$. After generating some examples we can take one in which the two new collisions have moved counterclockwise, resulting in high angular momentum, namely

\begin{equation}\label{segin2}\begin{split}
    X_{1,0} &= (0.3159, 0.6796, 0.7819, 0.1626)\\
    X_{0,1} &= (0.3254, 0.5087, 0.8648, 0.3152)\end{split}
\end{equation}

where we rounded the numbers just to be able to show them in a single line. When we rotate this by $\pi/8$, we retain the $\mathbb{Z}_{16}$ symmetry and have determined all $X_{i+1,-i}$. This initial data, as shown in Figure \ref{fig:exit1}, is all we need to evolve the string using \eqref{gubser}. From \eqref{segmass} and \eqref{segang} we find

\begin{equation}
    M = 5.90, \hspace{0.5cm} L = 5.97
\end{equation}

and therefore $M < L$. We expect that this implies that the string must leave the static patch, and indeed it can be seen that already at the next iteration half of the collisions lie outside the static patch. The string rotates counterclockwise and expands as it does so, as one can see in Figure \ref{fig:exit}. At the fourth step all collisions lie outside the static patch and finally, after the tenth step neighboring collisions are no longer causally connected. The evolution law breaks down and the string keeps expanding. As mentioned, the discrete collision points are all that is needed to determine the evolution, even without mentioning the string segments that connect them. It is pleasant however to see how the 2 kinks that emerge from each collision are connected by a fixed time curve in a totally geodesic $\dS_2$ fragment and collide again with other kinks. This is why we show in Figure \ref{fig:exit2} the string at times in between collisions, when one can distinguish 64 kinks.

\section{Strings captured by a black hole in \texorpdfstring{$\dS$}{TEXT}}\label{sec:bh}

The possibility of cosmic strings is an old subject \cite{Vilenkin:1984ib, Hindmarsh:1994re, Copeland:2003bj}. Recently though, interesting new phenomena have been discussed in \cite{Xing:2020ecz}. Their setup was to consider a string loop captured by a much more massive and compact black hole in four dimensions. At zeroth order it can be modelled, as a closed string in flat space, pinned to a specific point. The linearity of the equations of motion allows for a description using an auxiliary curve. Using this to their advantage, \cite{Xing:2020ecz} gave a beautiful and intuitive picture of the main effects in the evolution of such a string, being due to the black hole's finite mass,  torques acting on the black hole and possible superradiance effects. In $\dS$, as we have noticed, the equations become non-linear and much less pleasant. However, we can think of a situation where there is a (non)-rotating black hole with event horizon much smaller than the cosmic horizon, such that there is a regime close to the black hole where the evolution is very well described by \cite{Xing:2020ecz}, but needs to be complemented by curvature effects when it moves outwards too much.  

\subsection{Non-rotating black hole}
As argued in \cite{Xing:2020ecz}, in flat space, one can expect the string of length $s$ to develop self-intersections after a certain time. At reconnection, which for cosmic strings happens with probability $p \sim 1$ \cite{Vilenkin:2000jqa}, one expects the string to emit daughter loops, causing it to shrink. If $R$ is the Schwarzschild radius and $\mu$ the string tension, then for $p \gg \mu s/R$ 

\begin{equation}
    t_{\text{shrink}} = \frac{R}{\mu}.
\end{equation}

 The derivation starts from the periodicity of the auxiliary curve and therefore changes are expected. At intermediate radii however one expects the order of magnitude estimate to be correct. There is a second effect due to the torques of the moving string on the black hole, which is called horizon friction. The result is that the string loses energy and angular momentum to the black hole\footnote{This is not unlike the case of a string in $\AdS$, with one end moving at constant velocity on the boundary, and the other stretching down to the horizon. In such a configuration energy is also dissipated, resulting in the drag force calculated by \cite{Gubser:2006bz, Herzog:2006gh}.}, which  happens on a timescale

\begin{equation}
    t_\text{fr} = \frac{s^3}{R^2}.
\end{equation}

The derivation of this was rather general and only assumes that the tangent vectors to the string at the 2 points where it is attached to the black hole move on a timescale $s$. Again, non-linearity will become important, but at intermediate ranges such estimate is expected to hold. 

In $\dS$ as in flat space, the torques act as to decrease the energy and angular momentum of the string. In flat space this inevitably makes the string shorter. In $\dS$, we can consider the straight open string \cite{DeVega1996}, for which the energy and angular momentum both attain a maximum at \cite{Noumi2019}

\begin{equation}
    K[x]= 2 E[x], \hspace{0.5cm} x = \frac{1}{\omega^2 +1},
\end{equation}

with $E$ and $K$ the elliptic functions, corresponding to the previously mentioned $\omega \approx 0.4588$, see also Figure \ref{fig: blob}. When $\omega$ is smaller than this, the horizon friction will then tend to make the string longer until it stretches towards the horizon. 

For generic configurations, possible self-intersections and reconnections may quickly reduce $s$ until the flat space limit is reached. For 
\begin{equation}
s < \mu^{-1/3} R
\end{equation}

the horizon friction will be the most relevant effect, so for small enough $\mu$ one could expect the friction to be the most important one, even closer to the cosmic horizon. 

\subsection{Rotating black hole: defusing the bomb?}

In the rotating black hole there is, beyond the shrinking and friction, the additional interesting possibility of superradiance, where circularly polarized tension waves on the string get amplified upon reflection from the black hole. The fastest growth is achieved for modes of frequency 

\begin{equation}
    \omega = \frac12 \Omega_{BH} \cos{\theta_\text{min}}
\end{equation}

with $\theta_\text{min}$ being the minimum angle between the black hole spin axis and the string \cite{Xing:2020ecz}. In flat space, due to the periodicity of the string, after bouncing back, the tension wave will move along the string and  approach the black hole from the same direction once more. Modes are then amplified on a timescale of

\begin{equation}\label{bomb}
    t_{bomb} = \frac{s}{2(R\Omega_{BH}\cos{\theta_{\text{min}}})^2}.
\end{equation}

This leads to a black hole bomb scenario \cite{Press:1972zz}, where the string itself plays the role of a cavity \cite{Xing:2020ecz}. Additionally the length of the string increases. For a straight string in the flat space regime with $\Omega_{BH} \gg 1/s$, this happens as

\begin{equation}
    s = \sqrt{s^2_0 + 16\pi^2 R^2 \Omega_{BH}t}.
\end{equation}

Generically \eqref{bomb} with be the smallest timescale in the system and the exponential growth will manifest. For strings that are close enough to the $\dS_3$ equatorial plane it is possible that the lengthening process happens at a faster rate. When this happens and the string expands towards the horizon, the bomb will not go off. That excessive lengthening would spoil the linearity assumptions in deriving the black hole bomb process actually happens in flat space as well. The difference is that the $\dS$ horizon will have the additional effects of initially accelerating the expansion of the string and, as it moves close to the horizon, exponentially damping perturbations in static patch time.

In $\dS$, even before reaching the horizon the non-linearities in the equations of motion for the string will become important and the direction at which tension waves approach the black hole will no longer be strictly periodic, which would also disable the bomb mechanism, or at least reduce its efficiency.

The importance of reconnections in the rotating case, even in flat space, is not very clear, as \cite{Xing:2020ecz} already noted. It may well make a generic string stay within the flat space regime such that the above scenario where the string moves towards the horizon does not go through. In that case one could expect that already in flat space the bomb gets defused through reconnections. 

As noted in \cite{Xing:2020ecz} investigating this oscillatory behaviour and possible limit cycles through simulations deserves further attention. Considering the setup within the static patch through simulations is bound to make the phase space structure even more interesting, though arguably for realistic parameters its influence will be negligible. 

\section{Conclusion}

We showed that for rigidly rotating strings with center of mass at the origin of the static patch of $\dS_3$, with length scale $l$, the Higuchi bound $M l > L$ \eqref{higu} on energy $M$ versus angular momentum $L$ holds, as seen in Figure \ref{fig: higu2}, thereby extending the observation of \cite{Noumi2019} and providing evidence to the idea that all strings that remain within the static patch satisfy \eqref{higu}. These rotating strings are described by the Larsen-Sanchez class of solutions of the Polyakov action with constraints \cite{larsen1996}. We gave a brief review of these, with emphasis on the three types of string in Figure \ref{fig:types}. We argued that for strings with higher angular momentum per mass would end up expanding towards the horizon. 

After having commented on the unshifted Larsen-Sanchez solutions, leaving their relation to the classification \cite{bychen} for appendix \ref{app:cheng}, we approached the problem via the Nambu-Goto action. This proved useful in understanding uniqueness and perturbations, as well as for numerical analysis. In Figure \ref{fig:ngnumeric}, we gave a numerical example of a string that falls into the horizon and can therefore violate \eqref{higu}, zooming in on how its shape changes as it moves nearer to the horizon. We further discussed the case of the straight open rotor with point masses attached which always has a higher $Ml/L$ ratio than the corresponding open GKP-like string, as can be seen in Figure \ref{fig: blob}. 

We continued by approximating the classical string worldsheet in a piecewise manner by totally geodesic $\dS_2 \subset \dS_3$ segments, following the work done in $\AdS_3$ by \cite{Vegh:2015ska, Callebaut:2015fsa}. Those strings are exact solutions and can be described in terms of discrete lightcone coordinates, subject to the algebraic evolution equation of Gubser  \cite{Gubser:2016wno}. We derived discrete expressions for the energy and angular of the segmented string that are indeed invariant under this evolution. The algebraic evolution law breaks down when the expansion is sufficiently large for the kinks to end up in different causal patches, which typically happens some time after parts of the string left the static patch. 

For different initial conditions we generated random samples of segmented strings. The scatterplots in Figure \ref{fig:scatters} suggest that the probability for the string to remain in the static patch is mostly determined by its total energy. After that, we choose specific initial conditions with 64 kinks, and 32 kink collisions, resulting in a segmented string with $M l < L$. Whereas the evolution of the Nambu-Goto solution quickly became numerically imprecise, here we can very precisely follow how the string evolves in embedding space time, see Figure \ref{fig:exit}. The string leaves the static patch at the third iteration. The Gubser evolution law breaks down after the tenth iteration, when the collisions are no longer causally connected. After this the segmented string simply keeps expanding.

Finally we pointed out possible changes to the phenomena described by \cite{Xing:2020ecz} when considering their cosmic strings captured by black holes in case the configuration takes place in $\dS$ instead of flat space. For the case of a non-rotating black hole we found that the horizon friction can make the string longer. In this last section, particularly for rotating black holes, we were not able to provide much details, as the role of reconnections even in flat space is not well understood and calls for future research. It would be very interesting to figure out the complete phase space structure. 

Likewise it would be good to understand if \eqref{higu}, which was referred to as the Higuchi bound in \cite{ Noumi2019, Lust2019} has such a group theoretic interpretation at the level of classical strings. The Higuchi bound  \cite{Higuchi:1986py}, corresponds to positivity of the quadratic Casimir, for the principal and complementary series representations of $\SO(1, d+1)$. At large quantum numbers, taking all conserved charges zero, apart from the static patch energy and angular momentum, the quadratic Casimir reduces, at the classical level, to $M^2l^2- L^2$. For the rigidly rotating strings, which remain strictly within the static patch, this was shown to be positive. For strings that expand towards the horizon it can be negative, as we demonstrated in sections \ref{sec:ng11} and \ref{sec: segex}. This could be related to the fact that the string exits the horizon in finite proper time and that different parts of it lose causal contact after doing so. 

It would in general be interesting to understand better how strict the bound is for classical strings. For instance, is the value of $Ml / L$ for the open GKP-like string \cite{Noumi2019} a lower bound for all strings that remain in the static patch? 

Throughout this work, we assumed the backreaction on the geometry to be negligible. Moreover, we did not pay much attention to decay effects, of which there are essentially two types. The string could reconnect or split up at self-intersections, which will occur in the general case. This is a local effect and is not expected to change much in comparison with flat space. But even in the rigidly rotating case, the string will decay through massless emission, which is enhanced near the cusps \cite{Iengo2006}. Most likely, this will initially simply act to smooth out the cusps. In any case, all these effects disappear in the limit of vanishing string coupling. Gravitational self-interaction of the string \cite{Horowitz:1997jc, Damour:1999aw} could also be interesting in light of the correspondence with black holes.

We mostly discussed the 3-dimensional case as we can always restrict to strings lying in a $\dS_3$ subspace of $\dS_d$. Additionally this is also technically the more straightforward case to analyze. It could be fruitful to take a more general approach. Perhaps in this respect the many integrability results mentioned in the Introduction, such as the relation to the sinh-Gordon model via Pohlmeyer reduction \cite{pohlmeyer, DeVega1993}, or the band structure of the Lam\'e potential \cite{bakas2016}, could give valuable insights.\\

\textbf{\large Acknowledgements:} I would like to thank Nikolay Bobev, Frederik Denef, Kristof Dekimpe, Fri$\eth$rik Freyr Gautason, Yuri Levin, Jef Pauwels and Marco Scalisi for their useful suggestions and comments on the draft.

\newpage
\appendix

\section{Wick rotation of Kruczenski solutions}\label{app:Kruc}
The Kruczenski strings \cite{Kruczenski_2005} in  conformal gauge are given in \cite{jevicki2008}. By their Ansatz, they are rigidly rotating and upon Wick rotation from $\AdS_3$ to $\dS_3$ should therefore fit into the Larsen-Sanchez class. In $\AdS$ global coordinates we have

\begin{equation}
        r = \frac{l}{\sqrt{2}}(\cosh{2a}\; \text{cn}[u,k]^2+\cosh{2b}\; \text{sn}[u,k]^2-1)^{\frac12}
\end{equation}

with 

\begin{equation}
    u = (\frac{\cosh{2a}+\cosh{2b}}{\cosh{2a}-1})^{\frac12}\sigma, \hspace{0.5 cm} k = \frac{\cosh{2a}-\cosh{2b}}{\cosh{2a}+\cosh{2b}}.
\end{equation}

In $\AdS$ there are only 2 cases, with inward or outward cusps respectively, depending on whether $a < b$ or $a \geq b$. The well-studied case of spiky strings in $\AdS$ \cite{Kruczenski_2005} corresponds to the regime with outward cusps close to the boundary. These are dual to particular single-trace operators of the form 

\begin{equation}
    \mathcal{O} = \Tr{  D^k_+F  D^k_+F \cdots  D^k_+F}
\end{equation}

with $k$ large. The strings extend from $l\sinh{b}$ to $l\sinh{a}$. Shifting by a half-period gives complex results. We can rotate the above result by $l \mapsto \rmi l$ into $\dS$, correspondingly rotating $a$ and $b$,  and find

\begin{equation}
        r = \frac{l}{\sqrt{2}}(1-\cos{2a}\;\text{cn}[u,k]^2-\cos{2b}\; \text{sn}[u,k]^2)^{\frac12}
\end{equation}

with 

\begin{equation}
    u = \frac{\sigma}{l}(\frac{\cos{2a}+\cos{2b}}{1-\cos{2a}})^{\frac12}, \hspace{0.5 cm} k = \frac{\cos{2a}-\cos{2b}}{\cos{2a}+\cos{2b}}.
\end{equation}

For the regime with cusps, the above is real. With the previously used conventions for $c_1,c_2,k_1,k_2$ and making use of relation 8.169 of \cite{Gradshteyn:1702455} between the Weierstrass and Jacobi elliptic functions, we retrieve the previous expression for the Larsen-Sanchez solution upon identifying 

\begin{equation}
  \mathfrak{w}= \rmi \sqrt{2}\frac{\text{K}[1-k]}{\sqrt{\cos{2a}+\cos{2b}}}  .
\end{equation}

\section{Classification results of Chen}\label{app:cheng}

In \cite{larsen1996} it is noted that the conformal factor $\rme^\alpha$ of the worldsheet metric satisfies a sinh-Gordon, cosh-Gordon or Liouville type equation depending, in our notation, on the sign of $\cos{4a}- \cos{4b}$ (positive, negative or zero respectively). If $a=b$ or $\pi/2 - b$ it is zero and there is the Liouville type equation in lightcone coordinates

\begin{equation}
    \alpha_{+-} - \rme^\alpha = 0.
\end{equation}

This occurs precisely in the cases of constant Gauss curvature. Compact solutions within the static patch then degenerate to point particles moving at the speed of light. There are however the unshifted solutions that extend to the horizon describing more interesting minimal surfaces from the mathematical point of view. We can relate these to the ones we obtained via a classification result of \cite{bychen}.

There, the question is asked which Lorentzian minimal surfaces of constant Gauss curvature $1$ can occur in $\dS_{d+1}$. In other words, Lorentzian minimal surfaces for which the induced metric can be taken to have conformal factor $\alpha = -2 \log(u+v)+\log 2$. It is proven that one of three possibilities occurs. Either i) the surface is totally geodesic, or ii) it can be written as

\begin{equation}
    S(u,v) = \frac{z(u)}{u+v} - \frac{z'(u)}{2}
\end{equation}

for a certain curve $z$ in embedding space $\mathbb{R}^{1,d+1}$ such that


\begin{equation}
    (z,z) = 0, \quad (z',z')=4, \quad (z'',z'') = 0,\quad  z''' \neq 0
\end{equation}

with the embedding space inner product, or iii) it can be written as

\begin{equation}
     S(u,v) = \frac{z(u) + w(v)}{u+v} - \frac{z'(u)+w'(v)}{2}
\end{equation}

with more involved conditions on the curves $z$ and $w$, see \cite{bychen}. Here we will just look at some examples that can be obtained from i) and ii).

An example that gives i) is to take

\begin{equation}
    z(u) = (-2 - \frac{u^2}{2}, 2u,-2+ \frac{u^2}{2},0),
\end{equation}

$z''' = 0$ and therefore we have a totally geodesic $\dS_2 \subset \dS_3$. Indeed

\begin{equation}
     S(u,v) = (\frac{-4+uv}{2(u+v)}, -1 + \frac{2u}{u+v},-\frac{4+uv}{2(u+v)},0 )
\end{equation}

parametrizes this. Another simple example, of ii), is when we take as a seed curve

\begin{equation}\label{seed}
    z(u) = \sqrt{2}(\cosh u, \sinh u, \cos u , \sin u).
\end{equation}


The surface becomes

\begin{equation}
    S(u,v) = (\sqrt{2} \frac{\cosh u}{u+v}-\frac{\sinh u}{\sqrt{2}}, \sqrt{2} \frac{\sinh u}{u+v}-\frac{\cosh u}{\sqrt{2}}, \sqrt{2} \frac{\cos u}{u+v}+\frac{\sin u}{\sqrt{2}}, \sqrt{2} \frac{\sin u}{u+v}-\frac{\cos u}{\sqrt{2}}).
\end{equation}

Already here one sees that this surface is unbounded due to the denominator $u+v$. One can convert to static patch coordinates

\begin{equation}
    S(u,v) = r(u,v)( \cosh t(u,v), \sinh t(u,v), \cos \varphi(u,v) , \sin \varphi(u,v)).
\end{equation}

 Notably, this solution turns out to belong to the Larsen-Sanchez class, where the Weierstrass function is not shifted by the half-period, i.e. the ones that expand towards the horizon and indeed have infinite mass and angular momentum.  The above example has Larsen-Sanchez parameters $a = b= \pi/4$ and its simple expression allows to calculate the integrals for $M, L$ explicitly. Converting $u,v$ to $\tau = (u+v)/\sqrt{2}$ and $\sigma= (u-v)/\sqrt{2}$ we have
 
 \begin{equation}
     r = \sqrt{\frac12 + \frac{1}{\tau^2}}.
 \end{equation}
 
 We reach the horizon as $\tau \to \sqrt{2}$ and, when integrated up to $\tau$ we find
 
 \begin{equation}
     \frac{L}{M} = \frac{\tau \log|\sqrt{2}+\tau|-\tau \log|\sqrt{2}-\tau|-2\sqrt{2}}{\tau \log|\sqrt{2}+\tau|-\tau \log|\sqrt{2}-\tau|+2\sqrt{2}}
 \end{equation}
 
 both $M$ and $L$ diverge, but their ratio approaches one.



A seemingly different example can be found by taking as a seed curve 

\begin{equation}
    \sqrt{2}u^2 (\cosh{1/u}, \sinh{1/u}, \cos{1/u}, \sin{1/u}).
\end{equation}

However, it is the same solution as before, with coordinates $1/u$ and $1/v$. 

Other solutions can be found for instance by taking a static patch parametrization and a specific Ansatz for $\theta, \varphi$. Satisfying $(z',z')=4$ is done by taking 

\begin{equation}
r(u) = 2(\theta'^2 + \varphi'^2)^{-1/2}
\end{equation}

and then the condition on $(z'',z'')$ gives for each choice of $\varphi$ a nasty non-linear ODE for $\theta$. Generally, tractable Ans\"atze include the cases where $\varphi$ is a multiple of $\theta$. These turn out to give Larsen-Sanchez solutions once again, through conservation of misery. For example $\varphi = 2\theta$ results in a solution 

\begin{equation}
    \theta = 2 \frac{\sqrt{3}}{3} \arctan(\frac{u}{2\sqrt{3}}).
\end{equation}

Some trigonometric manipulations then lead to the conclusion that we again found ourselves a rigidly rotating solution, with $a= b= \pi/2 - \arctan2$. Taking instead $\theta = 2\varphi$ we get the same solution with $\tanh$ instead of $\tan$ and $a = \pi/2 - b = \arctan2$.


\newpage

\printbibliography

@article{DeVega1996,
author = {de Vega, H. J. and Egusquiza, I. L.},
doi = {10.1103/PhysRevD.54.7513},
issn = {15502368},
journal = {Physical Review D - Particles, Fields, Gravitation and Cosmology},
number = {12},
pages = {7513--7519},
title = {{Planetoid string solutions in 3+1 axisymmetric spacetimes}},
volume = {54},
year = {1996}
}

@article{Lust2019,
archivePrefix = {arXiv},
arxivId = {1907.04161},
author = {L{\"{u}}st, D. and Palti, E.},
doi = {10.1016/j.physletb.2019.135067},
eprint = {1907.04161},
issn = {03702693},
journal = {Physics Letters, Section B: Nuclear, Elementary Particle and High-Energy Physics},
pages = {7--9},
title = {{A note on string excitations and the Higuchi bound}},
volume = {799},
year = {2019}
}

@article{Noumi2019,
    author = "Noumi, T. and Takeuchi, T. and Zhou, S.",
    title = "{String Regge trajectory on de Sitter space and implications to inflation}",
    eprint = "1907.02535",
    archivePrefix = "arXiv",
    primaryClass = "hep-th",
    reportNumber = "KOBE-COSMO-19-12",
    doi = "10.1103/PhysRevD.102.126012",
    journal = "Phys. Rev. D",
    volume = "102",
    pages = "126012",
    year = "2020"
}

@article{DeVega1993,
author = {{De Vega}, H. J. and Sanchez, N.},
doi = {10.1103/PhysRevD.47.3394},
issn = {05562821},
journal = {Physical Review D},
number = {8},
pages = {3394--3404},
title = {{Exact integrability of strings in D-dimensional de Sitter spacetime}},
volume = {47},
year = {1993}
}

@article{Larsen1994,
archivePrefix = {arXiv},
arxivId = {hep-th/9303001},
author = {Larsen, A. L. and Frolov, V. P.},
doi = {10.1016/0550-3213(94)90425-1},
eprint = {9303001},
issn = {05503213},
journal = {Nuclear Physics, Section B},
number = {1-2},
pages = {129--146},
primaryClass = {hep-th},
title = {{Propagation of perturbations along strings}},
volume = {414},
year = {1994}
}

@article{Iengo2006,
archivePrefix = {arXiv},
arxivId = {hep-th/0601072},
author = {Iengo, R. and Russo, J. G.},
doi = {10.1088/1126-6708/2006/02/041},
eprint = {0601072},
issn = {10298479},
journal = {Journal of High Energy Physics},
volume = {2},
pages = {041},
primaryClass = {hep-th},
title = {{Handbook on string decay}},
year = {2006}
}

@article{DeVega1994a,
archivePrefix = {arXiv},
arxivId = {hep-th/9312115},
author = {de Vega, H. J. and Larsen, A. L. and S{\'{a}}nchez, N.},
doi = {10.1016/0550-3213(94)90643-2},
eprint = {9312115},
issn = {05503213},
journal = {Nuclear Physics, Section B},
number = {3},
pages = {643--668},
primaryClass = {hep-th},
title = {{Infinitely many strings in de Sitter spacetime: Expanding and oscillating elliptic function solutions}},
volume = {427},
year = {1994}
}

@article{larsen1996,
    author = "Larsen, A. L. and Sanchez, N. G.",
    title = "{Sinh-Gordon, cosh-Gordon and Liouville equations for strings and multistrings in constant curvature space-times}",
    eprint = "hep-th/9603049",
    archivePrefix = "arXiv",
    reportNumber = "ALBERTA-THY-08-96",
    doi = "10.1103/PhysRevD.54.2801",
    journal = "Phys. Rev. D",
    volume = "54",
    pages = "2801--2807",
    year = "1996"
}

@book{Gradshteyn:1702455,
      author        = "Gradshteyn, I. S. and Ryzhik, I. M. and
                       Zwillinger, D. and Moll, V.",
      title         = "{Table of integrals, series, and products; 8th ed.}",
      publisher     = "Academic Press",
      address       = "Amsterdam",
      month         = "09",
      year          = "2014",
      url           = "https://cds.cern.ch/record/1702455",
      doi           = "0123849330",
}

@article{Kruczenski_2005,
    author = "Kruczenski, M.",
    title = "{Spiky strings and single trace operators in gauge theories}",
    eprint = "hep-th/0410226",
    archivePrefix = "arXiv",
    doi = "10.1088/1126-6708/2005/08/014",
    journal = "JHEP",
    volume = "08",
    pages = "014",
    year = "2005"
}

@article{jevicki2008,
    author = "Jevicki, A. and Jin, K.",
    editor = "Fujiwara, Kazuhito and Itoyama, Hiroshi and Kawamoto, Shoichi and Kihara, Hironobu and Oota, Takeshi and Sakaguchi, Makoto and Takayanagi, Tadashi and Yasui, Yukinor",
    title = "{Solitons and AdS String Solutions}",
    eprint = "0804.0412",
    archivePrefix = "arXiv",
    primaryClass = "hep-th",
    doi = "10.1142/S0217751X0804113X",
    journal = "Int. J. Mod. Phys. A",
    volume = "23",
    pages = "2289--2298",
    year = "2008"
}

@article{gubser2002,
    author = "Gubser, S. S. and Klebanov, I. R. and Polyakov, Alexander M.",
    title = "{A Semiclassical limit of the gauge / string correspondence}",
    eprint = "hep-th/0204051",
    archivePrefix = "arXiv",
    reportNumber = "PUPT-2029",
    doi = "10.1016/S0550-3213(02)00373-5",
    journal = "Nucl. Phys. B",
    volume = "636",
    pages = "99--114",
    year = "2002"
}

@article{bakas2016,
title = "On elliptic string solutions in AdS3 and dS3",
journal = "J. High Energ. Phys.",
volume = "07",
year = "2016",
doi = "https://doi.org/10.1007/JHEP07(2016)070",
author = "I. Bakas and G. Pastras",
}

@article{pohlmeyer,
title = "Integrable Hamiltonian systems and interactions through quadratic constraints",
journal = "Commun.Math. Phys.",
volume = "46",
year = "1976",
pages = "207-221",
doi = "https://doi.org/10.1007/BF01609119",
author = "Pohlmeyer, K.",
}

@article{bychen,
author = {Chen, B.-Y.},
year = {2011},
month = {06},
pages = {485-503},
title = {Classification of minimal Lorentz surfaces in indefinite space forms with arbitrary codimension and arbitrary index},
volume = {78},
journal = {Publicationes mathematicae},
doi = {10.5486/PMD.2011.4860}
}

@article{Xing:2020ecz,
    author = "Xing, H. and Levin, Y. and Gruzinov, A. and Vilenkin, A.",
    title = "{Spinning black holes as cosmic string factories}",
    eprint = "2011.00654",
    archivePrefix = "arXiv",
    primaryClass = "astro-ph.HE",
    month = "11",
    year = "2020"
}

@article{Kruczenski:2006pk,
    author = "Kruczenski, M. and Russo, J. and Tseytlin, A. A.",
    title = "{Spiky strings and giant magnons on S**5}",
    eprint = "hep-th/0607044",
    archivePrefix = "arXiv",
    reportNumber = "IMPERIAL-TP-AT-6-5, PUPT-2203",
    doi = "10.1088/1126-6708/2006/10/002",
    journal = "JHEP",
    volume = "10",
    pages = "002",
    year = "2006"
}

@article{Kruczenski:2004kw,
    author = "Kruczenski, M. and Ryzhov, A. V. and Tseytlin, A. A.",
    title = "{Large spin limit of AdS(5) x S**5 string theory and low-energy expansion of ferromagnetic spin chains}",
    eprint = "hep-th/0403120",
    archivePrefix = "arXiv",
    reportNumber = "BRX-TH-537",
    doi = "10.1016/j.nuclphysb.2004.05.028",
    journal = "Nucl. Phys. B",
    volume = "692",
    pages = "3--49",
    year = "2004"
}

@article{Press:1972zz,
    author = "Press, W. H. and Teukolsky, S. A.",
    title = "{Floating Orbits, Superradiant Scattering and the Black-hole Bomb}",
    doi = "10.1038/238211a0",
    journal = "Nature",
    volume = "238",
    pages = "211--212",
    year = "1972"
}

@article{Lonsdale:1988xd,
    author = "Lonsdale, S. and Moss, I.",
    title = "{The Motion of Cosmic Strings Under Gravity}",
    doi = "10.1016/0550-3213(88)90003-X",
    journal = "Nucl. Phys. B",
    volume = "298",
    pages = "693--700",
    year = "1988"
}

@article{Frolov:1996xw,
    author = "Frolov, V. P. and Hendy, S. and De Villiers, J. P.",
    title = "{Rigidly rotating strings in stationary axisymmetric space-times}",
    eprint = "hep-th/9612199",
    archivePrefix = "arXiv",
    doi = "10.1088/0264-9381/14/5/015",
    journal = "Class. Quant. Grav.",
    volume = "14",
    pages = "1099--1114",
    year = "1997"
}

@article{Vilenkin:1984ib,
    author = "Vilenkin, A.",
    title = "{Cosmic Strings and Domain Walls}",
    reportNumber = "PRINT-84-0840 (TUFTS)",
    doi = "10.1016/0370-1573(85)90033-X",
    journal = "Phys. Rept.",
    volume = "121",
    pages = "263--315",
    year = "1985"
}

@article{Frolov:1988zn,
    author = "Frolov, V. P. and Skarzhinsky, V. and Zelnikov, A. and Heinrich, O.",
    title = "{Equilibrium Configurations of a Cosmic String Near a Rotating Black Hole}",
    reportNumber = "PRE-ZIAP-88-14",
    doi = "10.1016/0370-2693(89)91225-2",
    journal = "Phys. Lett. B",
    volume = "224",
    pages = "255--258",
    year = "1989"
}

@article{Frolov:1996be,
    author = "Frolov, V. P. and Hendy, S. and Larsen, A. L.",
    title = "{Stationary strings and principal Killing triads in (2+1) gravity}",
    eprint = "hep-th/9602033",
    archivePrefix = "arXiv",
    reportNumber = "ALBERTA-THY-04-96",
    doi = "10.1016/0550-3213(96)00140-X",
    journal = "Nucl. Phys. B",
    volume = "468",
    pages = "336--354",
    year = "1996"
}

@article{Hindmarsh:1994re,
    author = "Hindmarsh, M. B. and Kibble, T. W. B.",
    title = "{Cosmic strings}",
    eprint = "hep-ph/9411342",
    archivePrefix = "arXiv",
    reportNumber = "SUSX-TP-94-74, IMPERIAL-TP-94-95-5, NI-94025",
    doi = "10.1088/0034-4885/58/5/001",
    journal = "Rept. Prog. Phys.",
    volume = "58",
    pages = "477--562",
    year = "1995"
}

@article{Copeland:2003bj,
    author = "Copeland, E. J. and Myers, R. C. and Polchinski, J.",
    title = "{Cosmic F and D strings}",
    eprint = "hep-th/0312067",
    archivePrefix = "arXiv",
    doi = "10.1088/1126-6708/2004/06/013",
    journal = "JHEP",
    volume = "06",
    pages = "013",
    year = "2004"
}

@book{Vilenkin:2000jqa,
    author = "Vilenkin, A. and Shellard, E. P. S.",
    title = "{Cosmic Strings and Other Topological Defects}",
    isbn = "978-0-521-65476-0",
    publisher = "Cambridge University Press",
    month = "7",
    year = "2000"
}

@article{Higuchi:1986py,
    author = "Higuchi, A.",
    title = "{Forbidden Mass Range for Spin-2 Field Theory in De Sitter Space-time}",
    reportNumber = "YTP-86-06",
    doi = "10.1016/0550-3213(87)90691-2",
    journal = "Nucl. Phys. B",
    volume = "282",
    pages = "397--436",
    year = "1987"
}

@book{Frolov:1998wf,
    editor = "Frolov, V. P. and Novikov, I. D.",
    title = "{Black hole physics: Basic concepts and new developments}",
    doi = "10.1007/978-94-011-5139-9",
    volume = "96",
    year = "1998"
}

@article{Vegh:2015ska,
    author = "Vegh, D.",
    title = "{The broken string in anti-de Sitter space}",
    eprint = "1508.06637",
    archivePrefix = "arXiv",
    primaryClass = "hep-th",
    reportNumber = "CERN-PH-TH-2015-209",
    doi = "10.1007/JHEP02(2018)045",
    journal = "JHEP",
    volume = "02",
    pages = "045",
    year = "2018"
}

@article{Callebaut:2015fsa,
    author = "Callebaut, N. and Gubser, S. S. and Samberg, A. and Toldo, C.",
    title = "{Segmented strings in AdS$_{3}$}",
    eprint = "1508.07311",
    archivePrefix = "arXiv",
    primaryClass = "hep-th",
    reportNumber = "PUPT-2486",
    doi = "10.1007/JHEP11(2015)110",
    journal = "JHEP",
    volume = "11",
    pages = "110",
    year = "2015"
}

@article{Gubser:2016wno,
    author = "Gubser, S. S.",
    title = "{Evolution of segmented strings}",
    eprint = "1601.08209",
    archivePrefix = "arXiv",
    primaryClass = "hep-th",
    reportNumber = "PUPT-2489",
    doi = "10.1103/PhysRevD.94.106007",
    journal = "Phys. Rev. D",
    volume = "94",
    number = "10",
    pages = "106007",
    year = "2016"
}

@article{Scalisi:2019gfv,
    author = "Scalisi, M.",
    title = "{Inflation, Higher Spins and the Swampland}",
    eprint = "1912.04283",
    archivePrefix = "arXiv",
    primaryClass = "hep-th",
    doi = "10.1016/j.physletb.2020.135683",
    journal = "Phys. Lett. B",
    volume = "808",
    pages = "135683",
    year = "2020"
}

@article{Gubser:2006bz,
    author = "Gubser, S. S.",
    title = "{Drag force in AdS/CFT}",
    eprint = "hep-th/0605182",
    archivePrefix = "arXiv",
    reportNumber = "PUPT-2198",
    doi = "10.1103/PhysRevD.74.126005",
    journal = "Phys. Rev. D",
    volume = "74",
    pages = "126005",
    year = "2006"
}

@article{Herzog:2006gh,
    author = "Herzog, C. P. and Karch, A. and Kovtun, P. and Kozcaz, C. and Yaffe, L. G.",
    title = "{Energy loss of a heavy quark moving through N=4 supersymmetric Yang-Mills plasma}",
    eprint = "hep-th/0605158",
    archivePrefix = "arXiv",
    reportNumber = "NSF-KITP-06-36",
    doi = "10.1088/1126-6708/2006/07/013",
    journal = "JHEP",
    volume = "07",
    pages = "013",
    year = "2006"
}

@article{Kato:2021rdz,
    author = "Kato, M. and Nishii, K. and Noumi, T. and Takeuchi, T. and Zhou, S.",
    title = "{Spiky strings in de Sitter space}",
    eprint = "2102.09746",
    archivePrefix = "arXiv",
    primaryClass = "hep-th",
    reportNumber = "UT-Komaba/21-1, KOBE-COSMO-21-04",
    month = "2",
    year = "2021"
}

@article{Chu:2016pea,
    author = "Chu, C-S. and Giataganas, D.",
    title = "{Thermal bath in de Sitter space from holography}",
    eprint = "1608.07431",
    archivePrefix = "arXiv",
    primaryClass = "hep-th",
    reportNumber = "NCTS-TH-1607",
    doi = "10.1103/PhysRevD.96.026023",
    journal = "Phys. Rev. D",
    volume = "96",
    number = "2",
    pages = "026023",
    year = "2017"
}

@article{deVega:1994yz,
    author = "de Vega, H. J. and Larsen, A. L. and Sanchez, N. G.",
    title = "{Semiclassical quantization of circular strings in de Sitter and anti-de Sitter space-times}",
    eprint = "hep-th/9410219",
    archivePrefix = "arXiv",
    reportNumber = "DEMIRM-PARIS-94049",
    doi = "10.1103/PhysRevD.51.6917",
    journal = "Phys. Rev. D",
    volume = "51",
    pages = "6917--6928",
    year = "1995"
}

@article{Larsen:1995bp,
    author = "Larsen, A. L. and Sanchez, N. G.",
    title = "{New classes of exact multistring solutions in curved space-times}",
    eprint = "hep-th/9501101",
    archivePrefix = "arXiv",
    reportNumber = "DEMIRM-PARIS-95003, DEMIRM-OBS.-DE-PARIS-95003",
    doi = "10.1103/PhysRevD.51.6929",
    journal = "Phys. Rev. D",
    volume = "51",
    pages = "6929--6948",
    year = "1995"
}

@article{Horowitz:1997jc,
    author = "Horowitz, G. T. and Polchinski, J.",
    title = "{Selfgravitating fundamental strings}",
    eprint = "hep-th/9707170",
    archivePrefix = "arXiv",
    reportNumber = "NSF-ITP-97-097",
    doi = "10.1103/PhysRevD.57.2557",
    journal = "Phys. Rev. D",
    volume = "57",
    pages = "2557--2563",
    year = "1998"
}

@article{Damour:1999aw,
    author = "Damour, T. and Veneziano, G.",
    title = "{Selfgravitating fundamental strings and black holes}",
    eprint = "hep-th/9907030",
    archivePrefix = "arXiv",
    reportNumber = "IHES-P-99-54",
    doi = "10.1016/S0550-3213(99)00596-9",
    journal = "Nucl. Phys. B",
    volume = "568",
    pages = "93--119",
    year = "2000"
}

\end{document}